\documentclass[a4paper,fleqn,usenatbib]{mnras}
\usepackage[T1]{fontenc}
\usepackage{ae,aecompl}
\usepackage{graphicx}
\usepackage{amsmath}
\usepackage{amssymb}
\usepackage{mathtools}
\usepackage{multirow}
\usepackage{rotating}
\usepackage{xcolor}
\usepackage{url}
\usepackage{longtable}
\usepackage{hyperref}

\providecommand{\given}{\ensuremath{\hspace{0.01em}\mid\hspace{0.01em}}}

\providecommand{\gaia}{Gaia}
\providecommand{\gdr}[1]{GDR{#1}}

\providecommand{\gmag}{\ensuremath{G}}

\providecommand{\ra}{\ensuremath{\alpha}}
\providecommand{\dec}{\ensuremath{\delta}}
\providecommand{\glon}{\ensuremath{l}}
\providecommand{\glat}{\ensuremath{b}}
\providecommand{\sinb}{\ensuremath{\sin\glat}}
\providecommand{\pmra}{\ensuremath{\mu_{\ra\ast}}}
\providecommand{\pmdec}{\ensuremath{\mu_\dec}}
\providecommand{\bpg}{\ensuremath{{\rm bp\!-\!g}}}
\providecommand{\grp}{\ensuremath{{\rm g\!-\!rp}}}
\providecommand{\relvarg}{\ensuremath{{\rm relvarg}}}
\providecommand{\uwe}{\ensuremath{{\rm uwe}}}

\providecommand{\ncomp}{\ensuremath{Q}}
\providecommand{\feat}{\ensuremath{x}}
\providecommand{\prior}{\ensuremath{\pi}}
\providecommand{\info}{\ensuremath{{\cal I}}}
\providecommand{\pstar}{\ensuremath{P_{\rm star}}} 
\providecommand{\pqso}{\ensuremath{P_{\rm qso}}}
\providecommand{\pgal}{\ensuremath{P_{\rm gal}}}
\providecommand{\pext}{\ensuremath{P_{\rm ext}}} 
\providecommand{\expnqso}{\ensuremath{\widehat{N}_{\rm qso}}}
\providecommand{\expngal}{\ensuremath{\widehat{N}_{\rm gal}}}

\providecommand\mvec{{\ensuremath{\mathbf m}}}
\providecommand\tvec{{\ensuremath{\mathbf t}}}
\providecommand\Cvec{{\ensuremath{\mathbf C}}}
\providecommand\trans{{\ensuremath{^{\mathsf{T}}}}}

\providecommand{\degree}{\ensuremath{^\circ}}
\providecommand{\um}{\ensuremath{\mu\textrm{m}}}

\definecolor{VeryDarkBlue}{RGB}{0,0,80}
\definecolor{VeryDarkRed}{RGB}{90,0,00}
\hypersetup{ colorlinks = true, 
 urlcolor = VeryDarkBlue, 
 linkcolor = VeryDarkBlue,  
 citecolor = VeryDarkRed 
}

\title[Quasar and galaxy classification in \gaia\ DR2]{Quasar and galaxy classification in \gaia\ Data Release 2}

\author[C.A.L.\ Bailer-Jones et al.]{
Coryn A.L.\ Bailer-Jones, Morgan Fouesneau, Rene Andrae\\
Max Planck Institute for Astronomy, K\"onigstuhl 17, 69117 Heidelberg, Germany}

\date{Submitted 27 June 2019; Accepted 10 October 2019}

\pubyear{}

\begin{document}
\maketitle

\begin{abstract} 
  We construct a supervised classifier based on Gaussian Mixture Models to probabilistically classify objects in \gaia\ data release 2 (\gdr{2}) using only photometric and astrometric data in that release.  The model is trained empirically to classify objects into three classes -- star, quasar, galaxy -- for $\gmag \geq 14.5$\,mag down to the \gaia\ magnitude limit of $\gmag=21.0$\,mag.  Galaxies and quasars are identified for the training set by a cross-match to objects with spectroscopic classifications from the Sloan Digital Sky Survey.  Stars are defined directly from \gdr{2}.  When allowing for the expectation that quasars are 500 times rarer than stars, and galaxies 7500 times rarer than stars (the class imbalance problem), samples classified with a threshold probability of 0.5 are predicted to have purities of 0.43 for quasars and 0.28 for galaxies, and completenesses of 0.58 and 0.72 respectively. The purities can be increased up to 0.60 by adopting a higher threshold.   Not accounting for this expected low frequency of extragalactic objects (the class prior) would give both erroneously optimistic performance predictions and severely impure samples.  Applying our model to all 1.20 billion objects in \gdr{2} with the required features, we classify  2.3 million objects as quasars and 0.37 million objects as galaxies (with individual probabilities above 0.5).  The small number of galaxies is due to the strong bias of the satellite detection algorithm and on-ground data selection against extended objects. We infer the true number of quasars and galaxies -- as these classes are defined by our training set -- to be 690\,000 and 110\,000 respectively ($\pm 50$\%).
The aim of this work is to see how well extragalactic objects can be classified using only \gdr{2} data. Better classifications should be possible with the low resolution spectroscopy (BP/RP) planned for \gdr{3}.
\end{abstract}


\section{Introduction}
\label{sec:introduction}

The \gaia\ mission is designed to target stars in our Galaxy \citep{2016A&A...595A...1G}. Yet by recording all point-like objects brighter than some magnitude over the entire sky, it also observes many extragalactic objects. This is in fact desirable, because objects of effectively infinite distance are used to define the astrometric reference frame and to tie this to the International Celestial Reference Frame (ICRF)
\citep{2018A&A...616A..14G}.

The Gaia data processing system includes modules to classify and characterize extragalactic objects both using the onboard low-resolution BP/RP spectrophotometry and astrometry \citep{2008MNRAS.391.1838B,2013A&A...559A..74B,2018MNRAS.473.1785D,2012A&A...537A..42T}
and by image reconstruction \citep{2013A&A...556A.102K}
(see also \citealt{2015A&A...574A..46P}).
However, these results will only become available starting with the third \gaia\ data release planned for 2021. 

The second \gaia\ data release (hereafter \gdr{2}) provides, among other things, astrometry (positions, parallaxes, proper motions) and three-band photometry for up to 1.7 billion objects \citep{2018A&A...616A...1G}. There have been various efforts to identify extragalactic objects in the \gaia\ releases, mostly via a positional cross-match to other catalogues.
\cite{2019A&A...624A.145S}, for example, cross-matched the Sloan Digital Sky Survey (SDSS) Data Release 14 quasar catalogue to \gdr{2}, finding 399\,000 counterparts within 1\arcsec\ in the latter.
\cite{2019RAA....19...29L} compiled a list of 1.8 million quasars from various sources, about one million of which match to \gdr{2} objects to within 2\arcsec.
\cite{2015ApJS..221...12S} identified 1.4 million Active Galactic Nuclei (AGN) in the mid-infrared AllWISE catalogue using a two-colour criterion defined by \cite{2012MNRAS.426.3271M}.
The \cite{2018A&A...616A..14G} identified 557\,000 of these in a cross-match to \gdr{2}, which they used to put the \gdr{2} positions and proper motions on a global reference frame (called \gaia-CRF2) that is non-rotating relative to this sample. Their sample excludes detections within 5.7\degree\ of the Galactic plane, so the true number of AGN that \gaia\ can see will be higher.
\cite{2018ApJS..236...37P} cross-matched the same AGN sample from \cite{2015ApJS..221...12S} to \gdr{1}. This produced a sample of 568\,000 AGN for $\gmag<20.7$\,mag with an estimated 50\% completeness. They claim a stellar contamination of just 0.2\%, but as this is assessed from the fraction of their objects with SDSS spectra that are spectroscopically classified as stars, and SDSS spectroscopy is much less complete for stars than AGN,
this must be an underestimate of the true contamination.
\cite{2018RAA....18..118B} used colours from Pan-STARRS1 and AllWISE to construct a machine learning method to classify 80 million objects with counterparts in \gdr{2} as stars, quasars, or galaxies.

Numerous studies have been published that classify stars using \gdr{2} data (typically combined with non-\gdr{2} data), but there are very few published works that classify extragalactic objects using \gdr{2} data. \cite{2018A&A...615L...8H} looked for quasars by identifying objects with zero proper motion. When cross-matching these candidates to quasars colour-selected in WISE (the Wide-Field Infrared Survey Explorer), they conclude that within 1\degree\ of the north Galactic pole the contamination from stars for $\gmag=18$ to 20\,mag was around 25\%.  Other studies have looked for (and found) tens of lensed quasars \citep[e.g.][]{2018MNRAS.479.5060L, 2019A&A...622A.165D}.

In this paper we classify extragalactic objects in the \gdr{2} catalogue across the whole sky for $\gmag\geq 14.5$\,mag.  In order to classify as many of the objects as possible using a homogenous set of data, we only use features from \gdr{2}.  These include the colours, astrometry, and a measure of photometric variability. 
Several studies have shown that good quasar classifications can be obtained from multiband optical and/or infrared photometry \citep[e.g.][]{2015ApJS..219...39R,2015ApJS..221...12S,2019MNRAS.485.4539J}, so including additional photometry would improve our classifications for objects that have such data.
But beyond wanting to classify as much of \gdr{2} as possible, and doing so homogeneously, another motivation for this work is to determine how well classification can be done using a limited set of photometry and astrometry. The classifier we produce will contribute to the {\em Discrete Source Classifier} (DSC) in the \gaia\ data processing pipeline \citep{2013A&A...559A..74B}. DSC classifies \gaia\ sources primarily using the dispersed BP/RP spectra, but when these spectra are noisy, blended, or ambiguous, classifications based on other features can improve performance.

We classify objects probabilistically into the three classes {\em star}, {\em quasar}, and {\em galaxy}. Our model is trained empirically, using a subset of the \gdr{2} data, with extragalactic class labels provided externally from a cross-match to SDSS.  By not synthesising data we avoid many difficulties with selection functions, passbands, and zeropoints.
As \gdr{2} has no explicit morphological information (although we will exploit the fact that some of the features are sensitive to non-point-like character), our definition of the galaxy class is a subset of what we normally understand this class to be. The ``star'' class is essentially defined as the complement of these (albeit with some contamination). Our method accommodates class imbalance: the fact that extragalactic objects are much rarer than stars in \gdr{2}. This turns out to be very significant.

It should nonetheless be superseded by DSC's classifications of a significant fraction of the \gaia\ objects, based primarily on BP/RP spectra, to be published as part of \gdr{3} in the second half of 2021.

\section{Data}
\label{sec:data}

All of the input data to the model -- whether for the train, test, validation, or application phases --  are drawn from \gdr{2}.
Both the train and test sets are drawn from the same sample; in particular they have their true classes determined in the same way (explained below).
A validation set is one in which true classes have been determined by other authors from other data, and generally has a different feature distribution from the train/test set. An application set is one with unknown classes (here all of \gdr{2}).

We build a model only for objects with $\gmag \geq 14.5$\,mag, as we expect few brighter unresolved extragalactic objects.
The faint limit is $\gmag = 21.0$\,mag, set by the \gdr{2} catalogue.

\subsection{Classes}

We have three classes: stars, quasars, galaxies.
As we are using a supervised learning algorithm, the definition of our classes is determined by the training data, which we now discuss.

\subsection{Train and test data}

There are two ways of constructing a training set.
The first is synthetic, which uses models of the \gaia\ instruments to predict \gdr{2} data for simulated objects of known class. The second is empirical, in which we take \gdr2\ data for objects with classes determined from other surveys. We adopt the latter approach for a number of reasons:
\begin{enumerate}
\item It is difficult to get consistent object models for the wide range of types of stars, quasars, and galaxies that \gaia\ observes;
\item It is difficult to model the \gaia\ data accurately. This is particularly true for the (integrated) photometry;
\item It is difficult to model the \gdr{2} selection function, which arises not only from the survey, but also from the data processing and the numerous decisions made by the consortium of what to include in the catalogue.
\end{enumerate}
All of these difficulties would increase the misclassification rate, the magnitude of which would not be apparent from the test set.  By constructing instead an empirical training set from \gdr{2}, we are immune to many of these problems. We don't need to worry about selection functions, photometric passbands, astrometric zeropoints, or the inflation of uncertainties and so forth, because these are the same in the training, testing, and application sets.  What we do have to do, though, is assign classes reliably to a sufficient number of objects in \gdr{2} using information from other sources (which is not without its problems). We do this in the following way for each class.

\subsubsection{Quasars}\label{sec:data_train_quasars}
\begin{figure}
\begin{center}
\includegraphics[width=0.50\textwidth, angle=0]{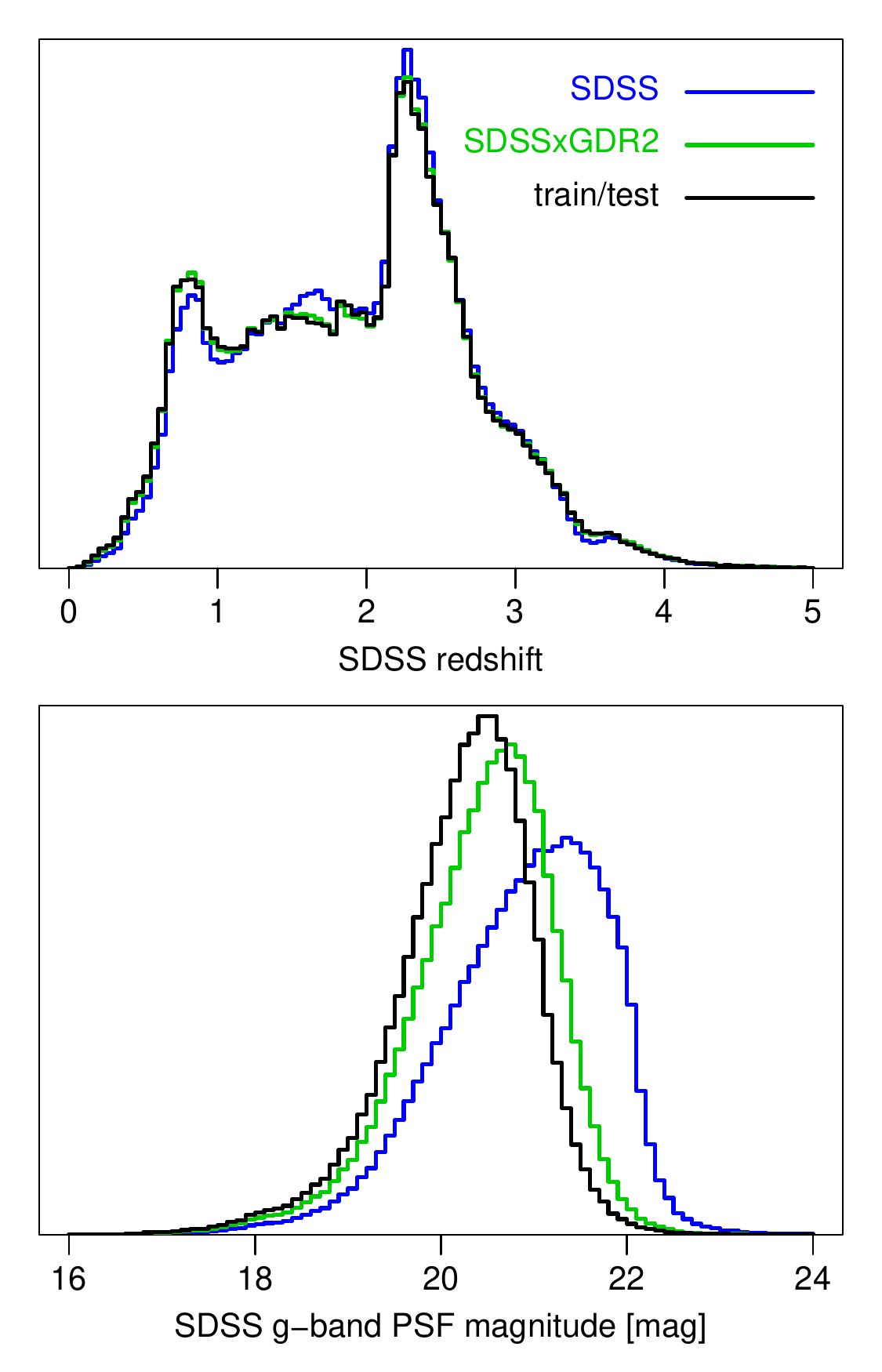}
\caption{The quasar redshift distributions (upper panel) and SDSS g-band magnitude distributions (lower panel) for the full SDSS sample (blue line, 446\,134 objects), what remains after the the cross-match with GDR2 (green line, 275\,940 objects) and what remains for our train/test set (i.e. have all 8 features; black line, 209\,287 objects).
Each distribution is separately normalized.
\label{fig:qso_redshift_mag_distributions}
}
\end{center}
\end{figure}

The SDSS-DR14 quasar catalogue contains 526\,356 quasars that have been confirmed by optical spectroscopy \citep{2018A&A...613A..51P}. Its authors estimate the contamination of this catalogue to be around 0.5\%.  We select those 446\,134 objects that have the {\tt zWarning} flag equal to zero, which indicates no issues with the classification or redshift determination.
We then cross-match these to \gdr{2} by sky position with a 1\arcsec\ search radius using the CDS X-match tool,
finding 275\,940 matches.  Of these, 209\,287 have the complete set of eight \gdr{2} features we use in our classifier (defined in section~\ref{sec:features}) and $\gmag \geq 14.5$\,mag. Figure~\ref{fig:qso_redshift_mag_distributions} shows the distribution of SDSS redshifts and magnitudes for the samples at each of the three phases of selection.  We see that requiring a cross-match to \gdr{2} (blue to green line) and then also requiring all eight features (green to black line) 
shifts the population to brighter magnitudes each time, but that the redshift distribution is hardly affected in either case.

Earlier in this project we experimented with using a ``quasar'' training set constructed from the
557\,000 AllWISE AGN cross-matched to \gdr{2} by \cite{2018A&A...616A..14G} (see section~\ref{sec:introduction}).
When we used these to define our ``quasar'' training set, we obtained comparable results to those we report below in terms of completeness and purity. However, a training set constructed from mid-infrared data will be a different population from that which \gaia\ sees. In particular, it will include dust-obscured quasars that are not be visible in the optical, leaving just the unobscured host galaxy visible to \gaia\ that would nonetheless have a ``quasar'' training label.

\subsubsection{Galaxies}\label{sec:data_train_galaxies}
\begin{figure}
\begin{center}
\includegraphics[width=0.50\textwidth, angle=0]{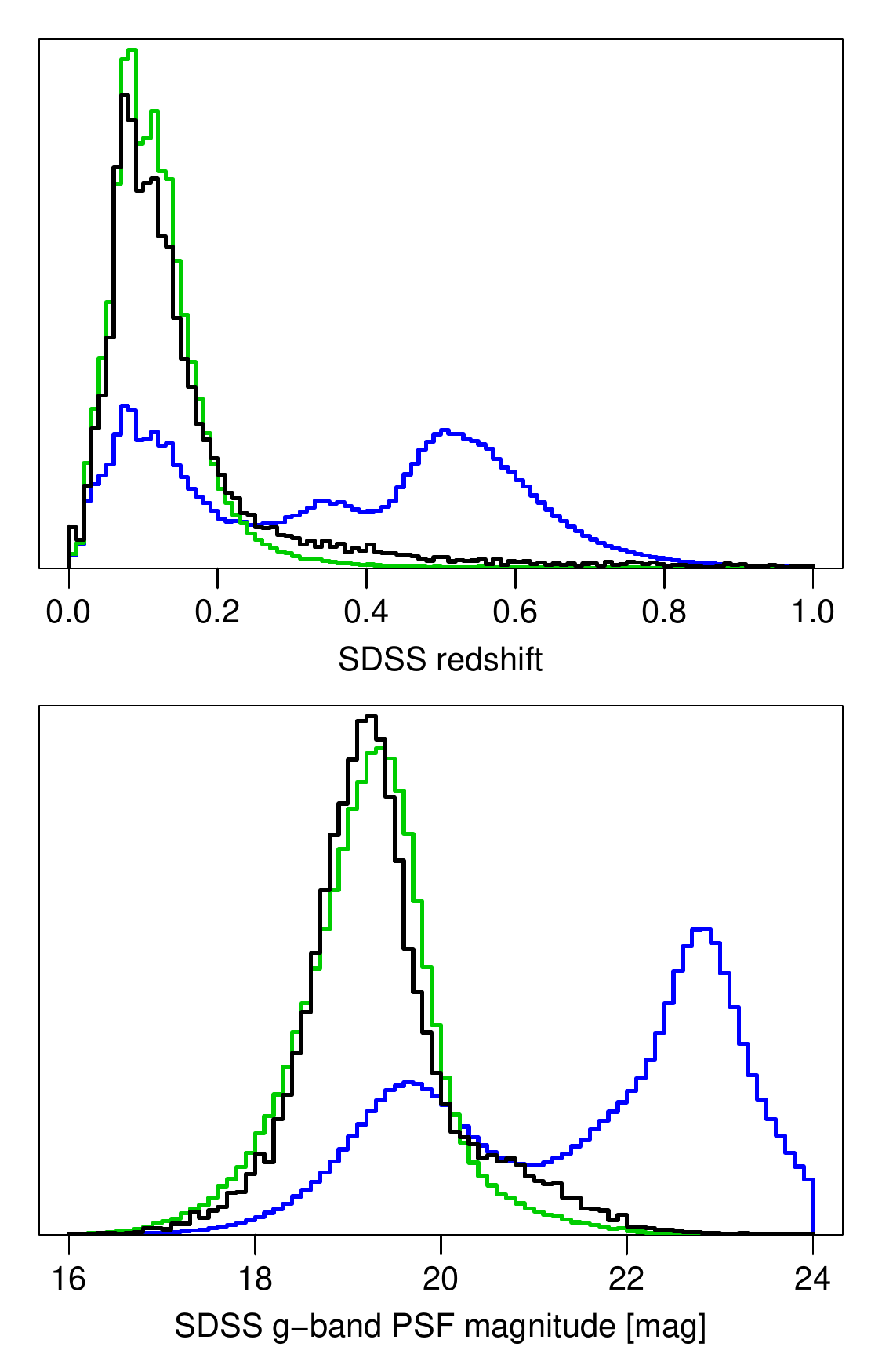}
\caption{The galaxy redshift distributions (upper panel) and SDSS g-band magnitude distributions (lower panel) for the full SDSS sample (blue line, 2\,303\,453 objects), what remains after the the cross-match with GDR2 (green line, 321\,965 objects) and what remains for our train/test set (i.e.\ have all 8 features and have $\grp \geq f(\bpg)$; black line, 18\,499 objects). Each distribution is separately normalized.
\label{fig:gal_redshift_mag_distributions}
}
\end{center}
\end{figure}

Our galaxy train/test set is constructed from SDSS-DR15 \citep{2019ApJS..240...23A}.  We select all objects from the {\tt SpecObjAll} table that have {\tt class} equal to {\tt GALAXY} and {\tt zWarning} equal to zero, and that do not have {\tt subClass} equal to either {\tt AGN} or {\tt AGN BROADLINE}.  This gives 2\,303\,453 objects. We then cross-match these to \gdr{2} in the same way as we did for the quasars, finding 321\,965 matches.  Of these, 19\,261 have the complete set of eight \gdr{2} features we need.
This number is much smaller than one might expect given \gaia's magnitude limit of $\gmag=21.0$\,mag. The reduction (by a factor of 7.2) in the number of objects when making the cross-match is much more than with the quasars (factor of 1.6), because \gaia\ only records objects with point-like image profiles, so most galaxies are excluded already on-board the satellite. The exact impact of the on-board detection algorithm on what is recorded is rather complicated \citep{2015A&A...576A..74D}.  Many galaxies that were recorded were then removed during the on-ground data processing, in particular if they gave poor solutions for the colours. The \gaia\ colours are computed from low resolution slitless spectrophotomtetry \citep{2018A&A...616A...4E}, so they can be corrupted by neighbouring sources. This is crudely quantified by the flux excess factor ({\tt phot\_bp\_rp\_excess\_factor}) in \gdr{2}, and all sources with this factor above 5 were removed from the catalogue \citep{2018A&A...616A...1G}.
Figure~\ref{fig:gal_redshift_mag_distributions} shows that there is a significant change in the redshift and magnitude distribution of the SDSS sample when we require the cross-match with \gdr{2} (blue to green line): we lose high redshift and faint galaxies.

Of the 321\,965 cross-matches we obtained, 94\% have 2-parameter astrometric solutions, i.e.\ just positions but no parallax or proper motions, and 28\% have no reported \gaia\ colours. The latter is almost entirely a subset of the former, i.e.\ almost everything that has no colour also has no parallax or proper motions.
This explains most of the large reduction (by a factor of 17) in the second selection step where we require the presence of all eight features.  Why so few galaxies have 5-parameter astrometric solutions, and why we nonetheless require this for our classifier, will be explained in section~\ref{sec:features} below.

\begin{figure}
\begin{center}
\includegraphics[width=0.50\textwidth, angle=0]{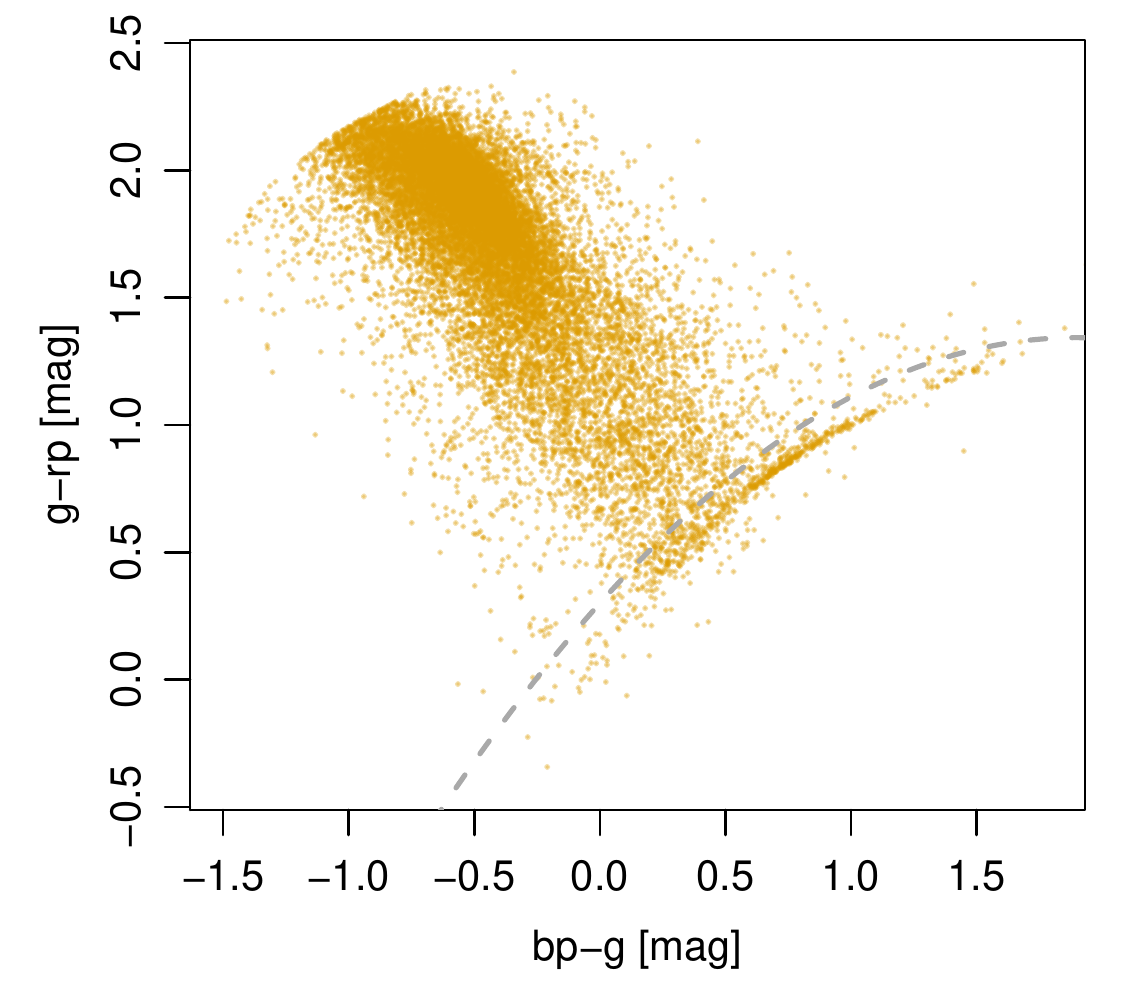}
\caption{Colour--colour diagram for the 19\,261 galaxies selected for the train/test set. The grey dashed line -- the ``colour edge'' -- is the locus defined by $\grp=f(\bpg)$ in equation~\ref{eqn:colouredge}. The 762 objects below this line are excluded from the train/test set.
  The smooth curved boundary at the top of the plot is a result of the cut in the flux excess factor imposed when \gdr{2} was constructed.
\label{fig:gal_all_withbtce_colcol}
}
\end{center}
\end{figure}

The \gaia\ colour-colour diagram of the remaining set of 19\,261 objects is shown in Figure~\ref{fig:gal_all_withbtce_colcol}.  In addition to the main bulk of galaxies in the upper left part of this plot, we see an additional, unexpectedly tight sequence of objects just below the grey dashed line. These are stellar contaminants, either errors in the SDSS classification, erroneous cross-matches to \gdr{2}, or BP/RP spectra corrupted by blends with nearby stars. We therefore define (by eye) the {\em colour edge}
\begin{equation}
  f(\bpg) = 0.3 + 1.1(\bpg) - 0.29(\bpg)^2
  \label{eqn:colouredge}
\end{equation}
shown by the grey dashed line in the plot, and remove all objects with $\grp < f(\bpg)$.
This leaves just 18\,499 galaxies in our train/test set (all have $\gmag \geq 14.5$\,mag). This is rather small, and is a reflection of the strong bias of the \gaia\ observing and data processing against retaining galaxies with 5-parameter astrometry.
We investigated supplementing this with other galaxy catalogues, in particular the 2dF galaxy redshift survey \citep{2001MNRAS.328.1039C}, the 6dF galaxy survey \citep{2009MNRAS.399..683J}, or the Galaxy And Mass Assembly (GAMA) survey \citep{2018MNRAS.474.3875B}. However, these did not add more than a couple of thousand galaxies. Using multiple surveys would also have made our definition of ``galaxy'' more heterogeneous, so we chose to use only the SDSS-DR15 sample.

Although requiring the eight features led to a significant reduction in the number of galaxies, this (together with the colour cut) has little impact on the redshift and magnitude distribution (green to black line in 
Figure~\ref{fig:gal_redshift_mag_distributions}): there is just a slight increase in the proportion of the faintest and most-redshifted galaxies. This indicates that the lack of parallaxes and proper motions is hardly correlated with the galaxy magnitudes and redshifts.

\subsubsection{Stars}
Building the star training set in the same way as the quasars and galaxies ones is unwise, because the SDSS spectroscopic selection function for stars is complex, poorly determined, and is unlikely to have a desirable distribution over stellar types.
We instead take advantage of the fact that, a priori, we expect most ($\geq99.8$\%) of \gdr{2} objects to be stars (see section~\ref{sec:prior_determination})
and take a random subset of all \gdr{2} objects to represent the ``star'' class. Although we remove known quasars and galaxies as defined by the previous classes, this set will still be contaminated. Exactly how contaminated we don't know, although the class prior (set in section~\ref{sec:prior_determination}) is the formal expectation. This contamination will inevitably reduce the accuracy of our classifications, albeit by a small amount here, and there is some hope that our model can generalize enough to be relatively insensitive to the contaminants. One must resist the temptation to identify the contaminants with the classifier, remove them from the training set, and then retrain the model, as this involves circular logic.

\subsubsection{Size of training and testing sets}

Our training set is a random subset of 100\,000 stars, 100\,000 quasars, and 10\,000 galaxies from the samples defined above.  The test is randomly constructed from the remaining objects, and comprises 100\,000 stars, 100\,000 quasars, and 8000 galaxies.  We will deal with the issue of class imbalance -- the different numbers of object in each class -- in section~\ref{sec:class_imbalance}.

\subsection{Validation data}\label{sec:data_validation}

We construct four validation sets in order to see how our model performs on data with very different properties and parameter distributions from our train/test sets. The first two of these we construct via a positional cross-match to \gdr{2}, for which we use a 1\arcsec\ radius. In all cases we retain only objects with all the eight features we need and $\gmag \geq 14.5$\,mag (these are the numbers given below)

The 2dF QSO Redshift Survey (2QZ) identified 23\,338 quasars spectroscopically, spread over two narrow bands on the sky \citep{2004MNRAS.349.1397C}. 
A cross-match of these to \gdr{2} left 18\,642 objects.

The 6dF galaxy survey (6dFGS) published spectra of 125\,071 galaxies spread over the southern sky \citep{2009MNRAS.399..683J}. A cross-match of those with a galaxy classification (\verb+S\G+=1) to \gdr{2} retained, after also removing any galaxies that lie beyond the colour edge (equation~\ref{eqn:colouredge}), 3017 objects.

\cite{2018ApJ...858L...7T} complied a set of 36\,909 ``pristine'' (their term) red clump (RC) stars identified using APOGEE spectra, for which they estimate the contamination from non-RC stars to be 3\% (their Table 2).
We identify these in \gdr{2} using the \gdr{2}-2MASS cross-match table provided in the \gaia\ archive.
10\,111 of the objects have our eight features and $\gmag \geq 14.5$\,mag. Most of them are at low Galactic latitudes.

The final validation set we use is a random subset of 500\,000 objects with distances below 500\,pc with 95\% confidence. These are drawn at random from
those 19 million objects with $r_{\rm hi}<500$\,pc in the distance catalogue of \cite{2018AJ....156...58B}. This is likely to be quite a pure set of stars, although there may be some contamination from anomalous astrometric solutions \citep{2018A&A...616A...2L}. We refer to this set as the ``nearby stars''.

\subsection{Application data}\label{sec:data_application}

The application data is the subset of \gdr{2} that has all eight features available and $\gmag \geq 14.5$\,mag.
There are 1\,203\,405\,908 such objects.

\subsection{Feature selection}\label{sec:features}

\begin{figure*}
\begin{center}
\includegraphics[width=1.0\textwidth, angle=0]{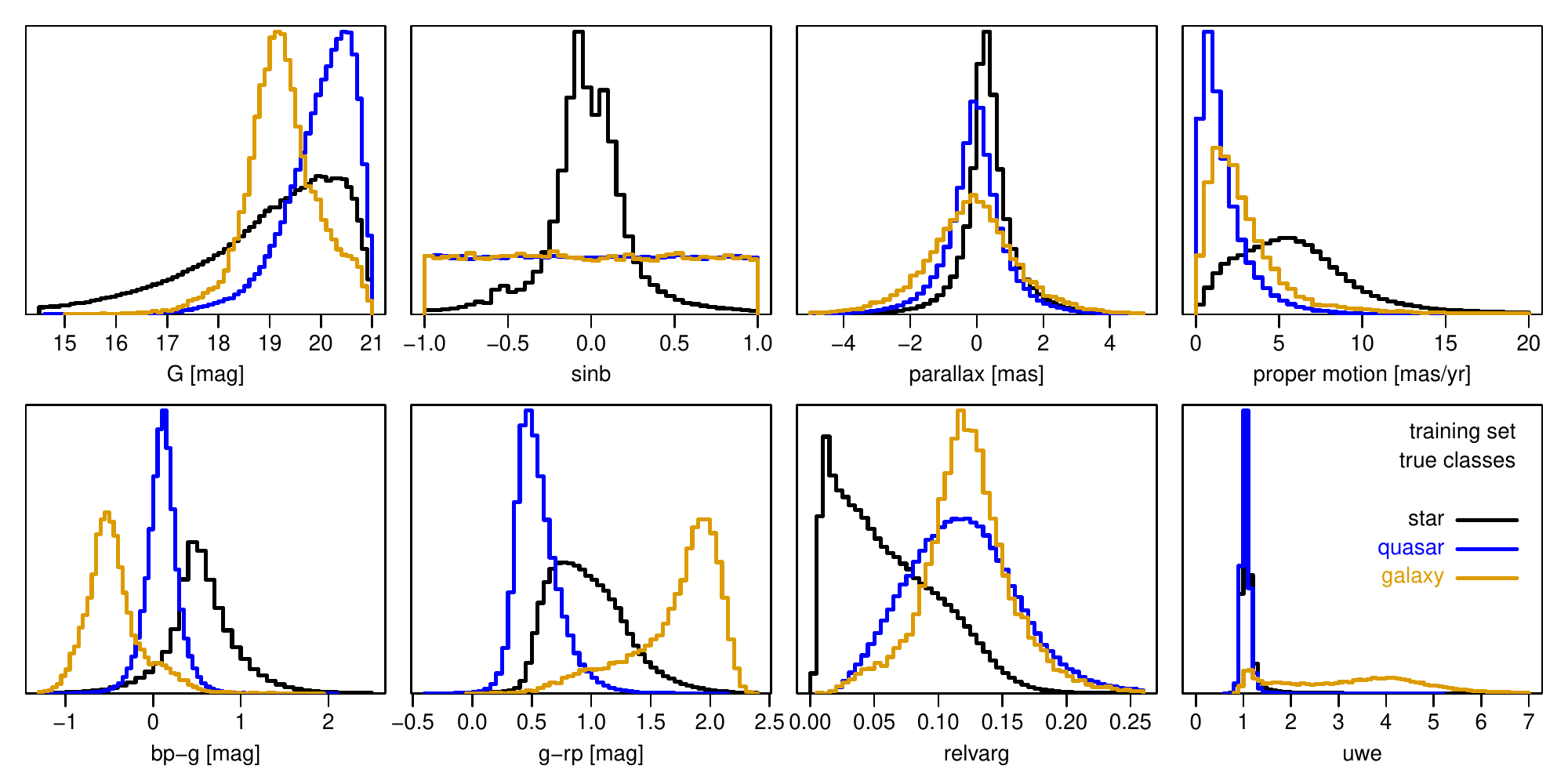}
\caption{The distribution of the eight features in the training set coloured according to their true classes.  Black: stars, blue: quasars, orange: galaxies. Each distribution is separately normalized. \sinb\ has been randomized for quasars and galaxies (constant probability per unit sky area).
\label{fig:any_all_featurehist}
}
\end{center}
\end{figure*}

\begin{figure}
\begin{center}
\includegraphics[width=0.50\textwidth, angle=0]{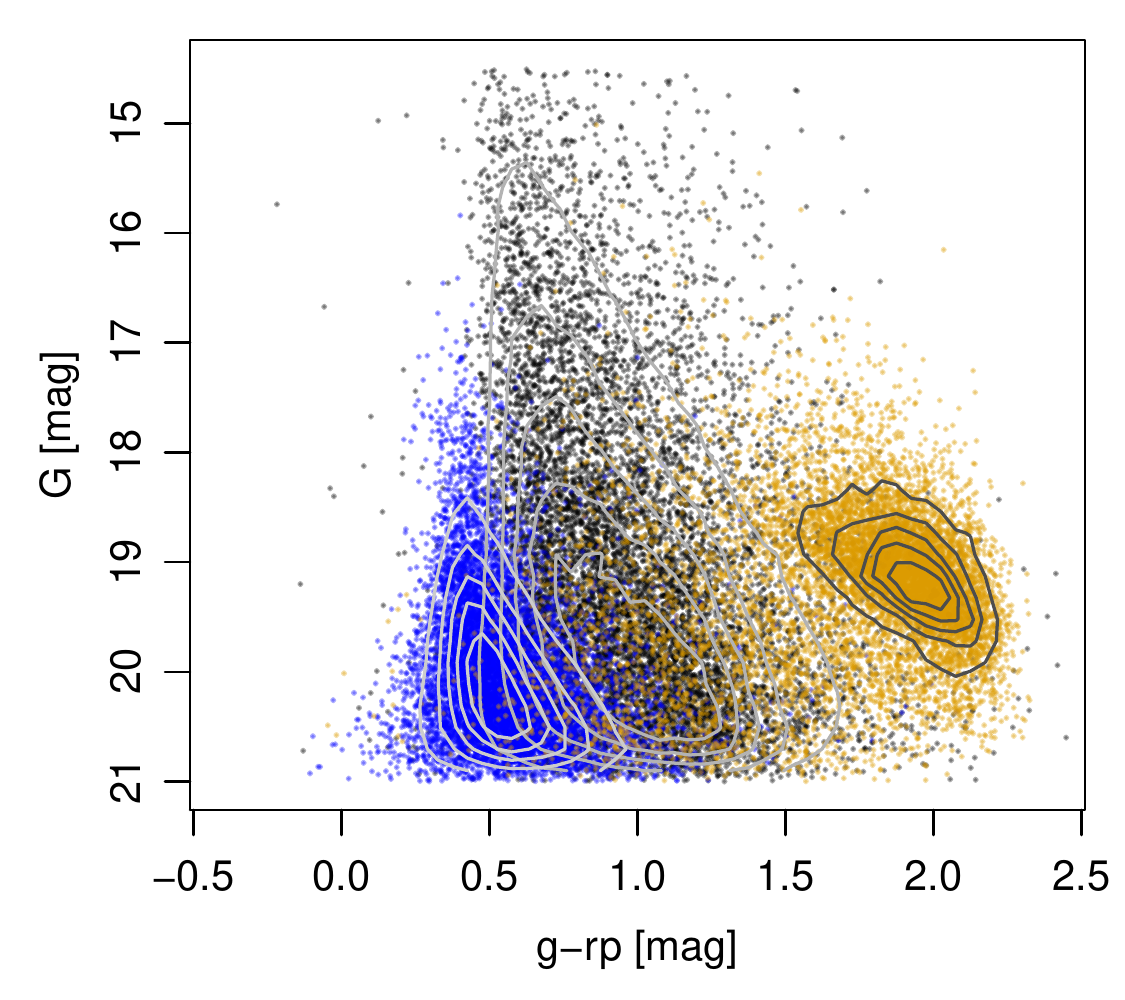}
\includegraphics[width=0.50\textwidth, angle=0]{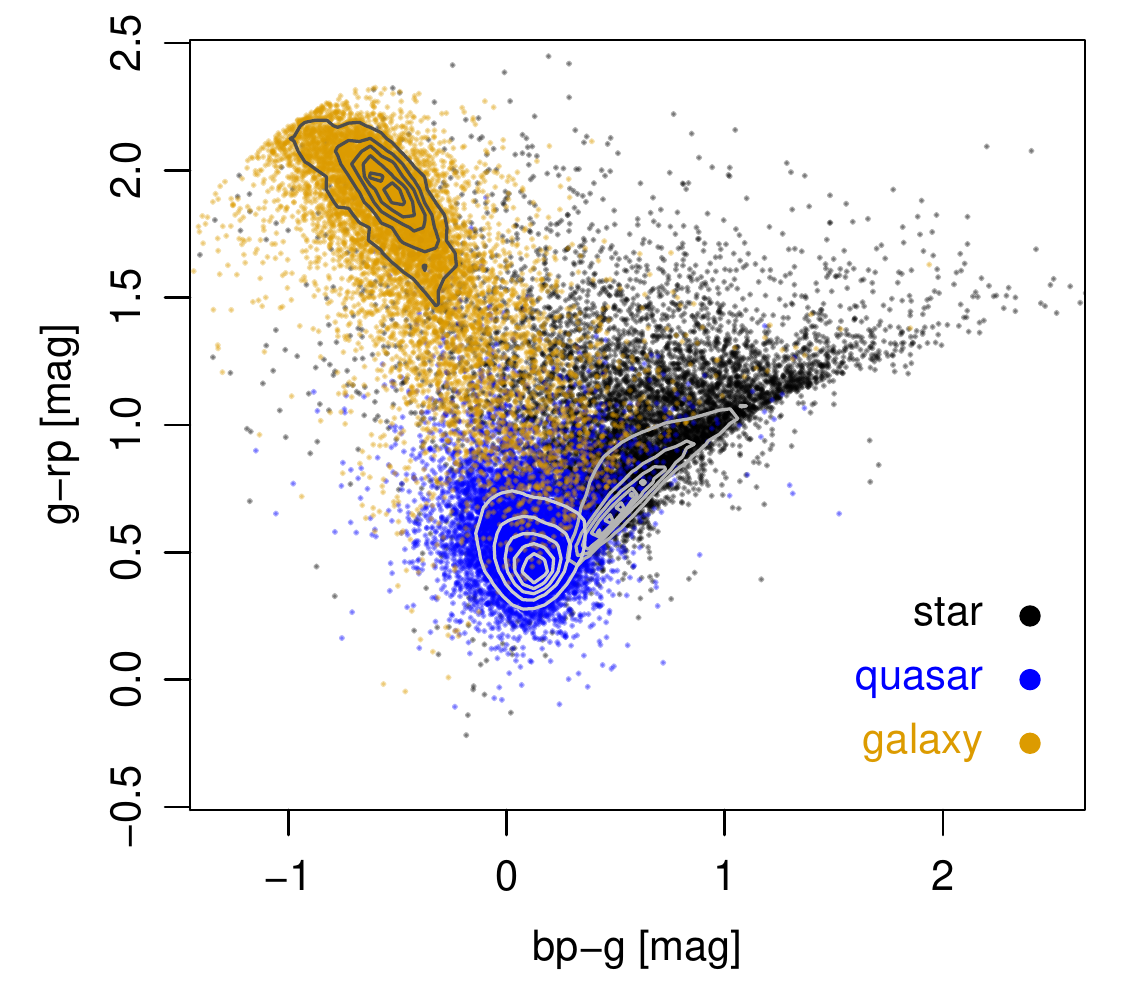}
\caption{Colour--magnitude diagram (top) and colour--colour diagram (bottom) for the training set coloured according to their true classes. Black: stars, blue: quasars, orange: galaxies (points plotted in that order).
    10\,000 objects have been selected at random from each class.
    The contours for each class (built from the entire train/test data set in each case) show equal spaced densities from
    $1/6$ to $5/6$ of the maximum density for that class.
The classes overlap, so we have plotted them separately in Figure~\ref{fig:any_plotsel_threepanels_onecmd} in the appendix.
\label{fig:any_plotsel_onecmd}
}
\end{center}
\end{figure}

We use a total of eight features in our models: apparent magnitude (\gmag), sine of the Galactic latitude (\sinb), parallax, total proper motion ($\sqrt{\pmra^2 + \pmdec^2}$), blue colour ($\bpg=\gmag-\gmag_{\rm BP}$), red colour ($\grp=\gmag_{\rm RP}-\gmag$), relative variability in the \gmag\ band (\relvarg), and the astrometric unit weight error (\uwe). These features we now define and discuss.

\gmag\ is the very broad pass band used by \gaia\ for doing the astrometry. Figure~\ref{fig:any_all_featurehist} shows that its distribution varies somewhat among the three classes, as does that of the two \gaia\ colours.
  This can also be seen in the colour--magnitude and colour--colour diagrams in Figure~\ref{fig:any_plotsel_onecmd}.
There is considerable class overlap, the impact of which will be exacerbated by the rareness of quasars and galaxies.

The motivation for using parallaxes and proper motions is the logical expectation that many stars have larger (more positive) parallaxes and proper motions than extragalactic objects, even in the presence of noise.  This is confirmed by our train/test sets, as we can see in Figure~\ref{fig:any_all_featurehist}.  We would expect the distributions of the proper motions of the quasars and galaxies to be the same, but they are in fact larger for galaxies. This
is because some galaxies are only just resolved by \gaia, and so different scans made at different position angles over the course of the mission reveal slightly different centroids.
This can be fit by the astrometric model as an erroneously large proper motion.  Something similar occurs for the parallaxes, where we would expect the distribution for both quasars and galaxies to be delta functions convolved with the noise, yet the distribution is distinctly wider for galaxies.

Galactic longitude and latitude $(\glon, \glat)$ are potentially useful, because whereas stars concentrate towards the Galactic disk and bulge, extragalactic objects should be more uniformly distributed. This uniformity is violated by our train/test data for quasars and galaxies, however,
because of SDSS's sky coverage. We therefore replaced the latitudes for these classes (in the train/test set) with values drawn from a uniform distribution in $\sinb$. (When applying the model to new data, we of course use the measured value of \glat.) This isn't entirely correct because, even if extragalactic objects cover the sky uniformly, our train/test sets should reflect the statistical properties of the classes as they appear in \gdr{2}, which is presumably a lower density at low $|\glat|$ due to dust absorption in the Milky Way disk. We simply accept this limitation.  While it may help us find otherwise-difficult-to-detect extragalactic objects at low latitudes, it may also lead to their higher false positive classification.

Galactic longitude is a problematic parameter because it wraps at $\glon=0\degree =360\degree$: If this is not known by the model, we see discontinuities in the spatial distributions of classified objects.  As wrapped parameters are awkward to deal with in a Gaussian Mixture Model (GMM), and including \glon\ didn't improve performance, we omit it.

We use two further features. The first is the fractional variability in the \gmag-band, the ratio of the standard deviation of the epoch photometry to its mean. (The timebase for \gdr{2} is 22 months.) We call this \relvarg, and it can be computed\footnote{$\relvarg = \frac{\sigma}{f} = \frac{\sqrt{N}}{(f/\sigma_s)}$, where $\sigma_s = \frac{\sigma}{\sqrt{N}}$ is defined in the catalogue as the error in the mean flux.} from the fields in \gdr2\ as $\sqrt{\verb#phot_g_n_obs#}/\verb#phot_g_mean_flux_over_error#$.
Figure~\ref{fig:any_all_featurehist} shows that quasars tend to be more variable than stars.
Galaxies too show large levels of variability, but as with the parallaxes and proper motions, this is largely a spurious result due to galaxie being extended: At each epoch scan, Gaia will determine a slightly different photocentre which can have different photometry. We can nonetheless exploit this behaviour to help distinguish galaxies.
 
The final feature is the unit weight error, \uwe. This is the square root of the $\chi^2$ per degree of freedom of the astrometric solution, computed from the fields in
\gdr{2} as $\sqrt{\verb#astrometric_chi2_al#/(\verb#astrometric_n_good_obs_al#-5)}$. Larger values generally indicate a poorer fit to the astrometric solution, and it shows an enhanced value for some galaxies (Figure~\ref{fig:any_all_featurehist}).

We investigated doing classification with a subset of these features. No one feature was critical. Performance degraded rather smoothly as features were removed.  We also experimented with using other features in \gdr{2}, such as the colour excess factor, but this did not improve performance. Other astrometric features such as the astrometric uncertainties or excess noise are quite correlated with \uwe.  Adding such features degraded performance (or did not improve it significantly), because they increase the dimensionality of the data space for little contribution of information.  There is not a large amount of relevant, independent classification information in \gdr{2}, and for a given number of training objects, lower-dimensional models are more robust than higher-dimensional ones.

\section{Methods}
\label{sec:methods}

\subsection{Classification model}

Many machine learning methods have been developed and used for supervised classification. Most of these, such as support vector machines (SVMs) and neural networks (NNs), model the boundary between classes in a multivariate data space, and use this to explicitly assign each object to one class \citep[e.g.][]{hastie}. However, rather than performing such a ``hard'' classification, we want to do {\em probabilistic} classification, i.e.\ determine the probabilities that each object belongs to all of the classes.  Not only does this give more flexibility in assigning classes (and thus adjusting the completeness and purity of samples, as we will see in later sections), it also provides us with a simple way of handling the crucial problem of class imbalance (in section~\ref{sec:class_imbalance}). Although one can get machine learning methods like SVMs and NNs to map their outputs to the range 0--1, this does not necessarily produce a reliable probability when the target labels were just 0s and 1s.

Probabilistic classification can be performed with a Gaussian Mixture Model \citep[GMM, e.g.][]{GMMs}. This models the distribution of the data in the $p$-dimensional feature space for each class using a sum of $K$ Gaussians. Specifically, the probability density for a feature vector \feat\ in class $C_k$ is 
\begin{equation}
P(\feat \given C_k) \,=\, \sum_{q=1}^\ncomp \, a_{k,q} \, \phi(\feat \given m_{k,q}, V_{k,q})
\label{eqn:likelihood}
\end{equation}
where $\phi$ indicates the density of a normalized multivariate Gaussian of mean $m$ and covariance matrix $V$.
The so-called mixing coefficients are subject to
$a_{k,q} \geq 0$, and to $\sum_{q=1}^\ncomp a_{k,q} = 1$ for all $k$.
A GMM with a sufficiently large number of components can in principle model any data distribution to arbitrary accuracy.
The means, covariance matrices, and mixing coefficients of the Gaussians are determined by maximizing the likelihood over the training data.
The GMM is then applied to a new object by computing the probability density for each class. When normalized, these values are the posterior class probabilities. This assumes equal class priors, but this will be relaxed (see section~\ref{sec:class_imbalance}).
A class is assigned by taking either the highest probability, or the highest probability that exceeds a threshold.
In the latter case, objects that do not exceed the threshold for any class are assigned to a $(K\!+\!1)$th class ``unclassified''.

For the GMM we use the \verb+mclust+ package (version 5.4 in R) of \cite{mclust2016}. 
The test data are not looked at during training.
Like other machine learning methods, the GMM contains many model parameters, of order $K\ncomp p(p+3)/2$.
This can be reduced by restricting the covariance structure of the Gaussians, for example by forcing all component Gaussians to be axis-aligned or to have the same shape. We experimented with this, but found that the best performance was obtained with no restrictions on the covariance structure (the so-called VVV model). The only parameter of the GMM is then the number of components \ncomp\ per class. We tested values between 5 and 30 and found that the performance
generally increased up until about 25 before levelling off. We therefore settled on $\ncomp=25$.

\subsection{Terminology}

The term {\em true class} cannot refer to some absolute truth, but rather to the assumed class, because the training, testing, and validation data inevitably have some labelling errors. The term {\em assigned class} refers to that which we assign using the probabilistic outputs of the classifier.
{\em Class fraction} is the fraction of objects of each class within a sample, so it is something we can measure given the classes.
{\em Class prior} is the class fraction we expect a sample to have, so is an assumption.

\subsection{Performance metrics}\label{sec:metrics}

We measure performance on the test set with the usual confusion matrix. Cell $i,j$ (row, column) of this is the number of objects of true class $i$ classified into class $j$. This is a $K \times K$ matrix if we classify by maximum posterior probability, and a $K \times (K+1)$ matrix if a threshold is used.
The former contains $K^2$ independent numbers, but the performance can be usefully summarized using two metrics for each class:\\
-- {\em completeness} (also called recall or sensitivity), the number of true positives as a fraction of all objects of that {\em true} class;\\
-- {\em purity} (also called precision), the number of true positives as a fraction of all objects of that {\em assigned class}.\\
We will mostly be interested in the completeness and purity of the quasar and galaxy classes, as these objects are rare and thus harder to classify.

\subsection{Handling class imbalance}\label{sec:class_imbalance}


A randomly selected sample of \gdr{2} objects will usually be dominated by stars, and thus have very unequal class fractions. This {\em class imbalance} must be accommodated both in the classifier design and in the reporting of results on a test set. Failure to do so gives both incorrect predicted performance on the test set and poor classification results on an application set. This can be seen by a simple example. Suppose we train a two-class classifier on the assumption of equal class fractions, and it gives a purity and a completeness of 0.9 in both classes on a balanced test set. If we now apply this classifier to an unbalanced data set
with $s$ times as many objects of class 1 as of class 2, then although the completenesses are both still 0.9, the purities are
$0.9s/(0.9s + 0.1)$ and $0.9/(0.1s+0.9)$ respectively.\footnote{The confusion matrix is
\begin{equation}
  \begin{pmatrix}
    0.9s & 0.1s \\
    0.1  & 0.9 \\
  \end{pmatrix}
  \nonumber
\end{equation}
with $s=1$ in training.}
For $s=100$, these purities are 0.9989 and 0.0826 respectively.
Class 1 dominates in this unbalanced set, so the class 2 output is dominated by misclassifications, making the purity much lower than that expected based on the balanced test set.

The initial reaction to this problem might be to train and test the model on appropriately imbalanced sets.
But this does not solve the problem, because a GMM models just the data distribution but not the amplitude in each class.  In fact, most machine learning methods do not learn the relative sizes of the training set classes in a way that would accommodate class imbalance (as shown, for example, by
\citealt{2008MNRAS.391.1838B}). Even if they did, then in order to get a sufficient number of training samples in the smallest class,
we would need $s$ times as many training samples in the larger class, which may not be available. Resampling already noisy observed data to achieve this is unwise because it increases the noise further.  Various ideas for dealing with this problem have been published in the machine learning literature \citep[e.g.\ SMOTE from][]{SMOTE2002}, but they are often complicated or cumbersome (e.g.\ by creating synthetic objects).

Fortunately, the problem is easily solved when using a probabilistic classifier like the GMM.
When applied to a new feature vector \feat, the GMM provides the likelihoods (equation~\ref{eqn:likelihood}) for classes $k=1 \ldots K$. These likelihoods are independent of the class balance in the training set, so we can train
with whatever number of objects is necessary in order to fit each class adequately.
Rather than just normalizing the likelihoods to give class probabilities (the equal prior case), we use Bayes' theorem to combine them with the class priors $\info = \{\prior_k\}$, to produce the posterior probabilities
  \begin{equation}
P(C_k \given \feat, \info)  \,=\, \frac{1}{Z} \, \prior_k \, P(\feat \given C_k)
\label{eqn:posterior}
\end{equation}
where $Z = \sum_k \prior_k P(\feat \given C_k)$.
As the posterior probabilities have changed, objects of a given true class (row of the confusion matrix) will be redistributed across assigned classes (columns of the confusion matrix). Thus {\em changing the class prior generally changes both the completeness and purity}.

In practice our class priors will never be equal to the true (unknown) class fractions, but any reasonable estimate (see section~\ref{sec:prior_determination}) will be far better than implicitly assuming equal class fractions. All results we report are from using posterior probabilities.

As this posterior classifier now assumes the data set has a particular class fraction, we will get a relevant confusion matrix only if the test set has the same class fraction.  If only a limited number of test objects are available, it seems we would have to cut down the number of objects in the smallest class in order to achieve the required class fraction. This would not only waste test data, it would also increase the noise in our performance metrics due to small-number statistics. Fortunately there is a simple remedy: We can use a balanced test set and then re-weight the confusion matrix to reflect the class prior, by multiplying each row of the confusion matrix by $\prior_k$. The numbers in this modified confusion matrix are no longer integers in general, but this is unimportant when we compute the completeness and purity.
This adjustment affects the purities (as that is a calculation across rows), but not the completenesses.

We can go one step further and adopt any size test set in each class that we like, e.g.\ to
avoid small number statistics when a classifier would otherwise assign very few objects to an output class. 
If the class fractions in this test set are $\{\alpha_k\}$, then
we multiply each row $k$ of the raw confusion matrix by the factor
\begin{equation}
\lambda_k \,=\, \frac{\prior_k}{\alpha_k} \left(\frac{1}{\sum_{k^\prime}(\prior_{k^\prime}/\alpha_{k^\prime})} \right) \ .
\label{eqn:testmodfac}
\end{equation}
This scales the actual number of objects in each row to the number of objects expected for a test set with class fractions $\{\prior_k\}$. 

\subsection{A random classifier} 

If, for a two-class problem, the target data set has class fractions $(0.99, 0.01)$, then it is easy to achieve a high completeness and purity on the first class (e.g.\ just assign everything to be class 1).
Thus at the minimum we should compare our performance to what a random classifier can do, defined as
a classifier that assigns classes at random according to the class priors.
For such a classifier the expected completeness and purity for class $k$ is $\prior_k$.
We will use this as a reference in section~\ref{sec:results}.

\section{Estimation of the class prior}\label{sec:prior_determination}

The relevant class prior is the expected class fraction in the sample of \gdr{2} data we will apply the classifier to. The only reliable way to determine these class fractions is if we had classifications for a representative subset of the data (here at least $10^6$ objects), but this does not exist. 

Part of the class prior we can hope to estimate to some degree of accuracy, however.  SDSS obtained spectra for as many galaxies and quasars as possible over a magnitude and wavelength range similar to \gaia's.  We can therefore use the values for the numbers of objects found in SDSS that cross-matched to \gdr{2} (together with the additional filtering) that we reported in sections~\ref{sec:data_train_quasars} and~\ref{sec:data_train_galaxies}. These were 209\,287 quasars and 18\,499 galaxies, a ratio of 11.3:1.  If we just look at the subset in stripe 82, which is a more systematic survey of what \gaia\ can see over a range of Galactic latitudes, the numbers are 6952 quasars and 402 galaxies, a ratio of 17.3:1. Neither of these values is a very accurate representation of the true, unknown, ratio in \gdr{2}, but it seems reasonable to adopt a class fraction of 15:1.

We cannot use SDSS in the same way to determine the ratio of stars to extragalactic objects, because stars were very selectively and sparsely targeted by SDSS. An early estimate during \gaia's study phase suggested that of one billion objects observed by \gaia\ down to $\gmag=20$\,mag, between one and ten million would be extragalactic \citep{2001A&A...369..339P}. The studies mentioned in section~\ref{sec:introduction} found up to one million quasars or AGN in \gdr{2}, which, among 1.7 billion sources, would imply a ratio to stars of 1700:1. However, these have certainly missed some quasars, especially at low Galactic latitudes. We choose to adopt a slightly higher ratio of 500:1.

We therefore adopt a class prior of (star, quasar, galaxy)\,=\,$(7500, 15, 1)$, which when normalized is $(0.997871, 0.001996, 0.000133)$.  This is probably accurate only to within a factor of a few.  While it will remain uncertain, it is the best we can do a priori, and is far better than assuming equal class priors. In section~\ref{sec:changing_prior} we investigate the impact on our results of varying this.

One of the benefits of our method is that it is simple to adjust the class probabilities to reflect a different prior. One simply divides the posterior probabilities we provide by the prior we used, multiply by the new prior, and renormalize (as used, for example, \citealt{2008MNRAS.391.1838B}).

\section{Results}
\label{sec:results}

Training the classifier takes around 20 minutes using a single CPU core. When applied to a new object, the classifier gives the posterior probabilities for the three classes (the probability vector). To be consistent with our definition of galaxies in section~\ref{sec:data_train_galaxies}, we must then make one modification. If an object has a colour beyond the colour edge of  equation~\ref{eqn:colouredge}, i.e.\ has $\grp < f(\bpg)$, it cannot be a galaxy, so we set $\pgal=0$ and renormalize the probability vector.

\subsection{Results on test data}\label{sec:results_test}


\begin{figure*}
\begin{center}
\includegraphics[width=1.0\textwidth, angle=0]{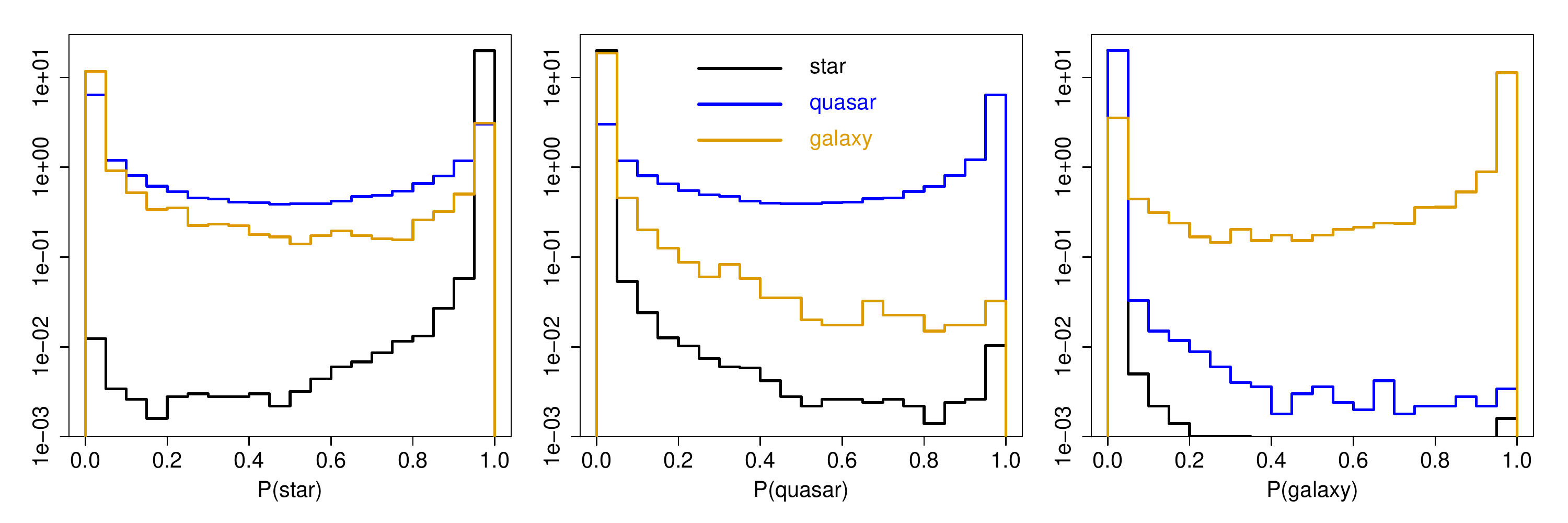}
\caption{The distribution of the three output probabilities (three panels) on the test set. The distributions for
  objects of the three true classes are shown in different colours (black: stars, blue: quasars, orange: galaxies).
  Each histogram in each panel is normalized. Note the logarithmic density scales. (The peak at low probabilities for stars in the right panel is under the one for quasars.)
\label{fig:any_test_posteriormodel_probhistlogscale}
}
\end{center}
\end{figure*}

The distributions of the posterior class probabilities obtained on the test set are shown in Figure~\ref{fig:any_test_posteriormodel_probhistlogscale}. This shows a peak towards high probabilities for the true class in all cases. For the quasars and galaxies there is also a lower peak at low probabilities for the true classes, indicating the loss of true objects due to overlap with the other classes in feature space. This plot is normalized separately for each class to allow an easier comparison of the probability distributions.  In reality, quasars and galaxies are much rarer than stars.

\subsubsection{Class assignment by maximum probability}\label{sec:results_test_maxprob}

\begin{table*}
\begin{center}
  \caption{Confusion matrix (and derived metrics) on the test set when assigning classes
    according to maximum posterior probability.
    The confusion matrix itself is the $3\times3$ subtable in the upper-left corner:
    It shows the numbers of objects we would get if the test set had the same class fraction as the class prior
    (which is why they are not integers; see section~\ref{sec:class_imbalance}).
    The completeness (compl.) and purity of each class derived from these numbers are shown in the next column and row.
    The following column and row marked ``(stdev)'' give the standard deviations in the completeness and purity over a set of 10 classifiers constructed with different randomly selected train/test sets. The penultimate column and row give, for comparison, the completeness and purity achieved if we adopt equal class priors in the classifier (with the same effective class fractions in the test as the main part of the table). The final column and row give the completeness and purity of a random classifier. 
\label{tab:cm_maxprob}
}
\begin{tabular}{c*{8}{r}}
\hline
&   & \multicolumn{3}{c}{assigned class} &           &             & equal prior: & random: \\ \cline{3-5}
&                    & star & quasar & galaxy & compl. & (stdev) & compl.         & compl. \\ \cline{3-9}
\parbox[t]{2mm}{\multirow{3}{*}{\rotatebox[origin=c]{90}{true class}}}
&  \multicolumn{1}{|r}{STAR}      & 99453 & 156 & 24.9 & 0.9982   & $\pm 0.0002$  & 0.9398 & 0.9979\\
&  \multicolumn{1}{|r}{QUASAR}  & 83.2 & 116 & 0.301  & 0.5811  & $\pm 0.0105$  & 0.9674 & 0.0020 \\
&  \multicolumn{1}{|r}{GALAXY}  & 3.46 & 0.164 & 9.67  & 0.7273  & $\pm 0.0184$  & 0.9343 & 0.0001  \\
  & purity                                   & 0.9991  & 0.4251 & 0.2771  & & \\
  & (stdev)                          & $<\pm 0.0001$ & $\pm 0.0164$ & $\pm 0.0351$ & & \\
& equal prior: purity                  & $>0.9999$ & 0.0504 & 0.0052  & & \\
& random: purity                       & 0.9979 & 0.0020 & 0.0001  & & \\
\hline
\end{tabular}
\end{center}
\end{table*}

We first assign classes by identifying the largest probability in the probability vector for each class. The resulting confusion matrix is shown in Table~\ref{tab:cm_maxprob}.\footnote{Recall that the confusion matrix 
  has been adjusted as explained in the final paragraph of section~\ref{sec:class_imbalance}. This adjustment only rescales each row; the ratios between the columns are unaffected. The raw confusion matrix is shown in Table~\ref{tab:cm_maxprob_raw}. If we row-normalize the confusion matrix (i.e.\ so that each row sums to one; this gives the same result for both confusion matrices) we get, in each row, the probabilities that a single object is classified into the three classes.} We achieve a purity of 43\% for quasars and 28\% for galaxies. These are not particularly high: The features available in \gdr{2} are not providing a very strong discrimination is the presence of many more stars.
  But these purities are not that bad when we consider that they assume that only every 1 in 500 objects is a quasar and only every 1 in 7500 is a galaxy.
The completenesses for both the quasars and galaxies are higher,  at around 58\% and 73\% respectively.
The purity on the galaxy class can vary by 3.5\% when we take different random selections of the train and test sets, as quantified by the standard deviation indicated in the table. This is another consequence of the rarity of galaxies,
as a small change in the modelling of the density of the star class can lead to a large change in the number of contaminants in the galaxy class.

Not using a class prior (or rather, assuming equal priors) in the classifier produces the completeness and purity shown in the penultimate column and row of Table~\ref{tab:cm_maxprob}.  The most striking result is how impure the quasar and galaxy samples become: 95\% of what we would classify as a quasar, and 99.5\% of what we would classify as a galaxy, would be incorrect. This is because such a classifier doesn't know that these classes are rare, so it assigns too many objects to them. Most of these are of the wrong class, so the purity is low.  Because this equal-prior classifier is too liberal in classifying things as quasars and galaxies, it also captures more of them, and so the completeness of these classes actually increases. But this cannot justify not using a prior, as a sample with such large contamination is nearly useless. (We discuss this further in section~\ref{sec:changing_prior} below.)



\subsubsection{Class assignment by threshold}\label{sec:apply_assigment_by_threshold}

\begin{figure}
\begin{center}
  \includegraphics[width=0.50\textwidth, angle=0]{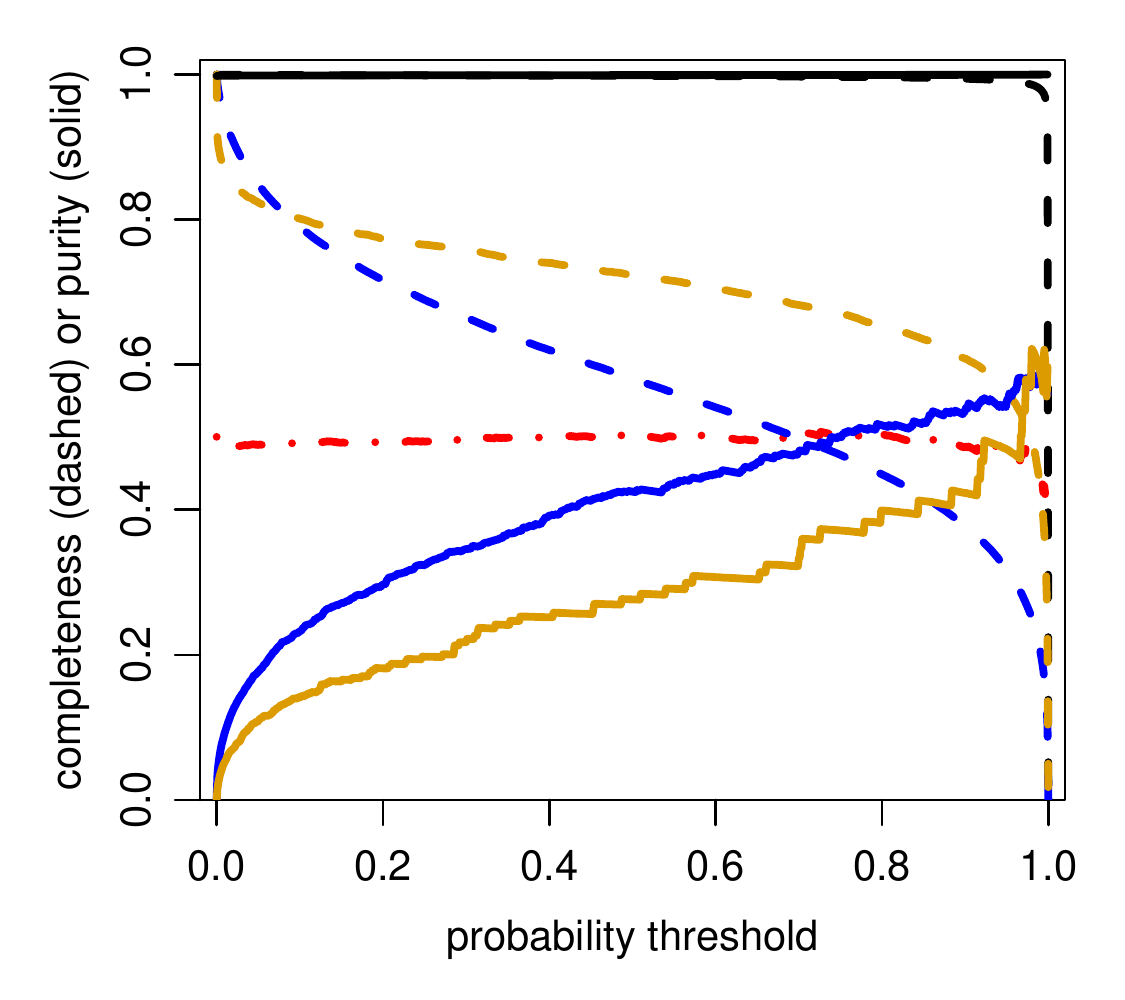}
  \includegraphics[width=0.50\textwidth, angle=0]{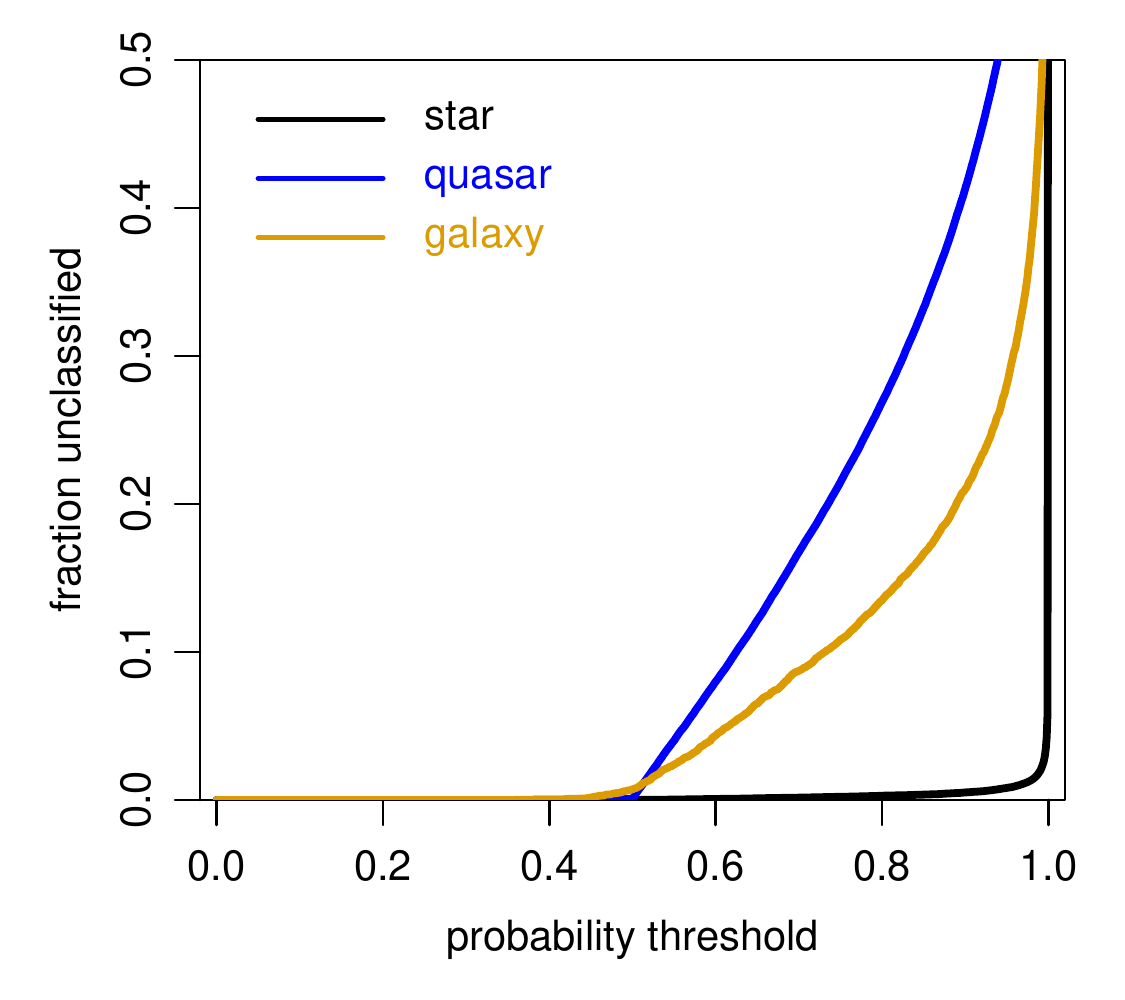}
\caption{Top: The variation of the completeness (dashed lines) and purity (solid lines) of each class (black: stars, blue: quasars, orange: galaxies) with the posterior probability threshold (same for all classes) used for assigning classes. This is computed on the test set. The red dot-dash line shows the average of the completeness and purity for the quasar and galaxy classes.
  Bottom: The corresponding variation in the fraction of objects in each true class that remain unclassified. The vertical axis in this panel only extends to 0.5 (where the probability is well above 0.999 for stars, 0.939 for quasars, and 0.993 for galaxies)
  but the lines continue up to 1.0.
\label{fig:any_testmod_rptrend}
}
\end{center}
\end{figure}

We can instead choose to assign objects to classes only if the posterior probability exceeds a threshold.  The higher the threshold, the higher the purity of the resulting sample, but the lower its completeness.  This is shown in the upper panel of Figure~\ref{fig:any_testmod_rptrend}. If we took a threshold of, say, 0.8 on the quasar output, then we would achieve a completeness of 0.45 and a purity of 0.52. The same threshold for galaxies would give a completeness of 0.65 and a purity of 0.40. If we consider the sum of the completeness and purity for the quasar and galaxy classes as a single metric (the red dot-dash line in the upper panel of Figure~\ref{fig:any_testmod_rptrend}), then we see that a broad range of thresholds provides similar performance (although we might be able to do better, as we are free to use different thresholds on different classes).

Although a completeness of 1.0 can be achieved trivially by setting the threshold to zero, there is a
limit to what purity can be achieved, even at very high thresholds. This is about 0.60 for both quasars and galaxies.

\begin{table*}
\begin{center}
\caption{As Table~\ref{tab:cm_maxprob}, but now assigning classes only if the posterior probability exceeds 0.5 (upper table) or 0.8 (lower table). ``Unclassified'' gives the (effective) number not classified into any class (as they have probabilities under the threshold), for each true class.
\label{tab:cm_threshold}
}
\begin{tabular}{c*{6}{r}}
\hline
& threshold=0.5   & \multicolumn{4}{c}{assigned class} & \\ \cline{3-6}
&                    & star & quasar & galaxy & unclassified & compl.\ \\  \cline{3-7}
\parbox[t]{2mm}{\multirow{3}{*}{\rotatebox[origin=c]{90}{true class}}}
&  \multicolumn{1}{|r}{STAR}       & 99452 & 156 & 24.9 & 0.996 & 0.9982 \\
&  \multicolumn{1}{|r}{QUASAR}  & 83.1 & 116 & 0.267 & 0.151 &  0.5809 \\
&  \multicolumn{1}{|r}{GALAXY}   & 3.43 & 0.143& 9.62 & 0.0963 & 0.7240 \\
& purity                                    &  0.9991 & 0.4250  &  0.2764 & \\
\end{tabular}
\begin{tabular}{c*{6}{r}}
\hline
&  threshold=0.8 & \multicolumn{4}{c}{assigned class} & \\ \cline{3-6}
&                    & star & quasar & galaxy & unclassified & compl.\ \\  \cline{3-7}
\parbox[t]{2mm}{\multirow{3}{*}{\rotatebox[origin=c]{90}{true class}}}
&  \multicolumn{1}{|r}{STAR}       & 99250 & 83.7 & 13.0 & 288  & 0.9961 \\ 
&  \multicolumn{1}{|r}{QUASAR}  &  56.2 & 89.4 & 0.106 & 53.6 & 0.4488 \\
&  \multicolumn{1}{|r}{GALAXY}   & 2.77 & 0.0548 & 8.67 & 1.79 & 0.6528 \\
  & purity                                    & 0.9994  & 0.5164 &  0.3991 & \\
\hline
\end{tabular}
\end{center}
\end{table*}

The price we pay when assigning thresholds is that a given object can either be assigned to more than one class (if the thresholds are low enough, e.g.\ all less than 0.5) or assigned to no class at all (if the thresholds are high enough) and so remain unclassified.

Table~\ref{tab:cm_threshold} shows the confusion matrices for thresholds of 0.5 and 0.8 (same on all classes).  For 0.5 the completenesses and purities are very similar to those from taking the maximum probability (cf.\ Table~\ref{tab:cm_maxprob}), which makes sense when we see that just 0.001\% of stars, 0.08\% of quasars, and 0.7\% of galaxies go unclassified. 
Raising the thresholds to 0.8 increases these to 0.3\%, 27\%, and 13\% respectively.  The lower panel of Figure~\ref{fig:any_testmod_rptrend} shows how the fraction of unclassified objects varies with the threshold.

\subsubsection{Changing the class prior}\label{sec:changing_prior}


The class prior is not well constrained (section~\ref{sec:prior_determination}), so what is the effect on the performance if we change it from the adopted value of (star, quasar, galaxy)\,=\,$(7500,15,1)$?  We do not need to refit the GMM to do this, so it is easy to modify our classifications to reflect a different prior.
Recall that when we change the class prior we not only modify the model output probabilities, but also the 
(effective) class fraction in the test set (as explained in section~\ref{sec:class_imbalance}). This must be done to retain consistency of the meaning of the class prior.

If we increase the star:quasar ratio by a factor of two, but keep the quasar:galaxy ratio constant, so the class prior is now $(15\,000,15,1$), the purities for quasars and galaxies decrease to 0.31 and 0.20 respectively. The completenesses are also lowered, to 0.51 and 0.70 respectively. Increasing the star:quasar ratio by another factor of two, so quasars are now 2000 times rarer than stars (and galaxies 30\,000 times rarer), the purities for quasars and galaxies become 0.21 and 0.14 respectively, and the completenesses 0.45 and 0.66. Clearly, if extragalactic objects were this rare, we would need more discriminating features (i.e.\ spectra) to classify them with an acceptable purity.

We now keep the star:quasar ratio constant, but increase the quasar:galaxy ratio to 30:1, so the class prior is now $(7500,15,0.5$).  Quasar performance is virtually unchanged from the nominal results (in Table~\ref{tab:cm_maxprob}), but the galaxy completeness drops by 0.03 to 0.69 and the purity drops by 0.09 to 0.19. If, conversely, galaxies were twice as common as in the nominal class prior, i.e.\ $(7500,15,2$), then galaxy completeness and purity instead increase by the similar amounts, to 0.75 and 0.38 respectively.

The closer we push the class prior towards equal priors, the better the measured performance.  Does this imply we should simply use equal priors? No, because the class prior is not arbitrary: It is our estimate of the class fractions in the application set. That is why we must also modify the (effective) class fraction in the test set (as explained in section~\ref{sec:class_imbalance}), i.e.\ it must be representative of the application set.
If we use equal class priors (and with a balanced test set), then we get purities of 0.94 for the quasars and 0.96 for the galaxies, and completenesses of 0.97 and 0.93 respectively. This looks good, but in the real application set, both quasars and galaxies are much rarer. So if we applied this equal-prior model to the real (unbalanced) application set, we would get {\em much} lower purities of 0.050 for quasars and 0.0052 for galaxies (these are shown in Table~\ref{tab:cm_maxprob}).

This is an important conclusion: If we overrepresent a class in the prior, the performance will be predicted to be better that that achievable with the correct prior, yet the actual performance would be not only be worse than this prediction, it would be worse than the performance obtained with the correct prior. It is essential to not only use an appropriate class prior in the classifier, but also to estimate the performance using a test set which has (effective) class fractions close to what we expect.

There is, however, one use for setting the class priors equal in the above experiment: It gives us a measure of the ``intrinsic'' ability of the GMM to distinguish between the classes, independently of class balance issues.  Doing this we get completenesses of 0.94, 0.97, 0.93 and purities of 0.94, 0.95, 0.96 for stars, quasars, and galaxies respectively. This shows that it is not intrinsically difficult to get high purities on both quasars and galaxies.  The reason we are expecting much lower values on \gdr{2} (i.e.\ Table~\ref{tab:cm_maxprob}) is because of heavy contamination by the much more numerous stars.

\subsection{Results on validation data}

We now apply our model to the validation sets described in section~\ref{sec:data_validation}.
These have neither the parameter distributions of our training data set nor the class fractions that our model is expecting, so the significance of these results should not be over-interpreted.

We first classify the quasars in 2QZ. 15\,223 objects -- a fraction of 0.82 of the total -- are classified as quasars (by maximum posterior probability).  This compares to an expected completeness of 0.58  (Table~\ref{tab:cm_maxprob}).  The fact that 2QZ gives a higher completeness could just indicate that the feature distribution of 2QZ is tightly clustered on the higher probability regions for quasars that made our train/test set in Figure~\ref{fig:any_all_featurehist}. In fact, the distributions of all the features are quite similar, except for Galactic latitude: the 2QZ sample is located entirely at high latitudes, an indication to the model of lower star probability.

For objects in 6dFGS, only 42\% are classified as galaxies by our classifier, compared to an expected completeness of 73\%. The reason is the converse of the one just given for quasars: the 6dFGS galaxies (that match to \gdr{2}) have quite a different distribution from our training set. In particular, they extend redder in \bpg\ and have far more sources that are bluer in \grp. They are also generally brighter and have an excess of objects with very low \relvarg, which is absent in our training sample. This demonstrates the caution one must exercise when assessing performance on independent validation sets.

We next apply our model to the red clump stars. Our model classifies all of them as stars with very high confidence. The largest value of \pqso\ is $4\times 10^{-6}$ and of \pgal\ is $5\times 10^{-5}$. Ting et al.\ estimate their sample to have a 3\% contamination, but this is presumably by other stars rather than quasars or galaxies.  

Finally we apply our model to the 500\,000 objects in the ``nearby stars'' validation set. 99.83\% are classified as stars, the same as the expected completeness.
193 objects are classified as quasars and 643 as galaxies. 
Based on the confusion matrix in Table~\ref{tab:cm_maxprob}, we would have expected 157 to be misclassified as quasars and 25 as galaxies. The larger number of galaxies is probably because these nearby stars have larger proper motions than most stars, and are so more similar to the (spuriously-large) proper motions of the galaxies.

\subsection{Results on application data (all of \gdr{2})}\label{sec:results_application}


1\,203\,405\,908 of the 1.69 billion objects in \gdr{2} have all eight features and $\gmag \geq 14.5$\,mag.  Applying our classifier to these using a single CPU core took 17 hours.  We find 2\,287\,566 objects with $\pqso>0.5$ and 367\,555 objects with $\pgal>0.5$.  If we instead classify according to the maximum probability, we get 2\,297\,133 quasars and 378\,219 galaxies.

Defining the extragalactic probability as $\pext=\pqso+\pgal$, we find 2\,690\,021 objects with 
$\pext>0.5$. We call this the {\em extragalactic sample}.
All other objects of course have $\pstar\geq0.5$ and so are classified as stars.

%

\subsubsection{Distribution in feature space }\label{sec:results_apply_feature_space}

\begin{figure*}
\begin{center}
\includegraphics[width=1.0\textwidth, angle=0]{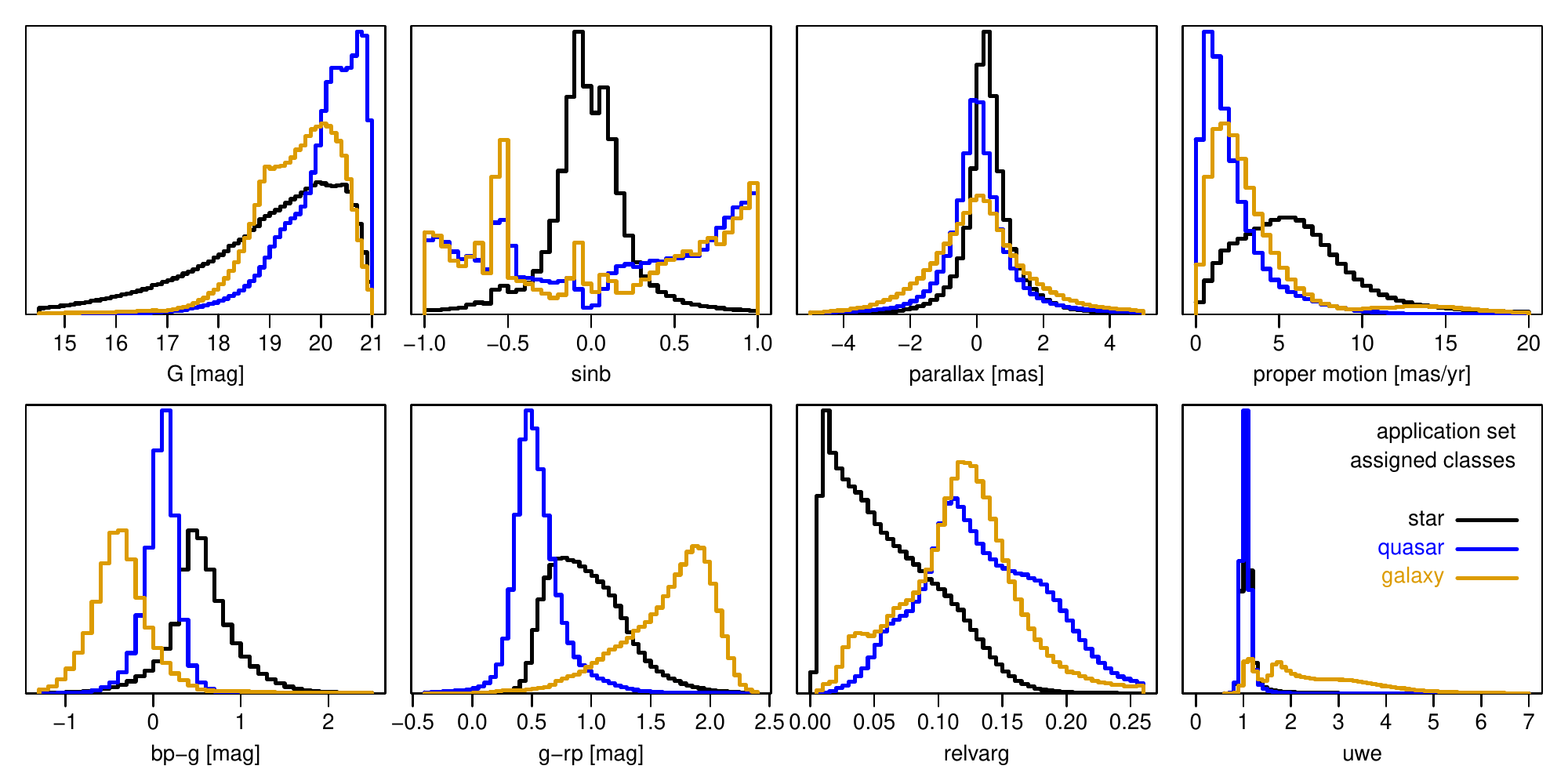}
\caption{The distribution of the eight features for objects in the application set assigned to classes when the posterior probability is greater than 0.5. The colour indicates the assigned class, black: stars, blue: quasars, orange: galaxies.
  Each distribution is separately normalized.
\label{fig:apply_any_0p5threshold_featurehist}
}
\end{center}
\end{figure*}

The distributions of the features in the application set for objects with posterior probabilities greater than 0.5 are shown in Figure~\ref{fig:apply_any_0p5threshold_featurehist}. These are colour-coded by the {\em assigned} class, and as the quasars and galaxies are predicted to have a contamination of more than 50\% (see Table~\ref{tab:cm_threshold}), care must be taken when comparing them to the features colour-coded by the {\em true} class in Figure~\ref{fig:any_all_featurehist}.

We see that the distributions for the parallax, proper motion, and the two colours are quite similar for test and application for all three classes.  Hence, most stellar contamination in the two extragalactic classes occurs from that subset of stars which have a very similar distribution in these parameters.

\begin{figure}
\begin{center}
\includegraphics[width=0.50\textwidth, angle=0]{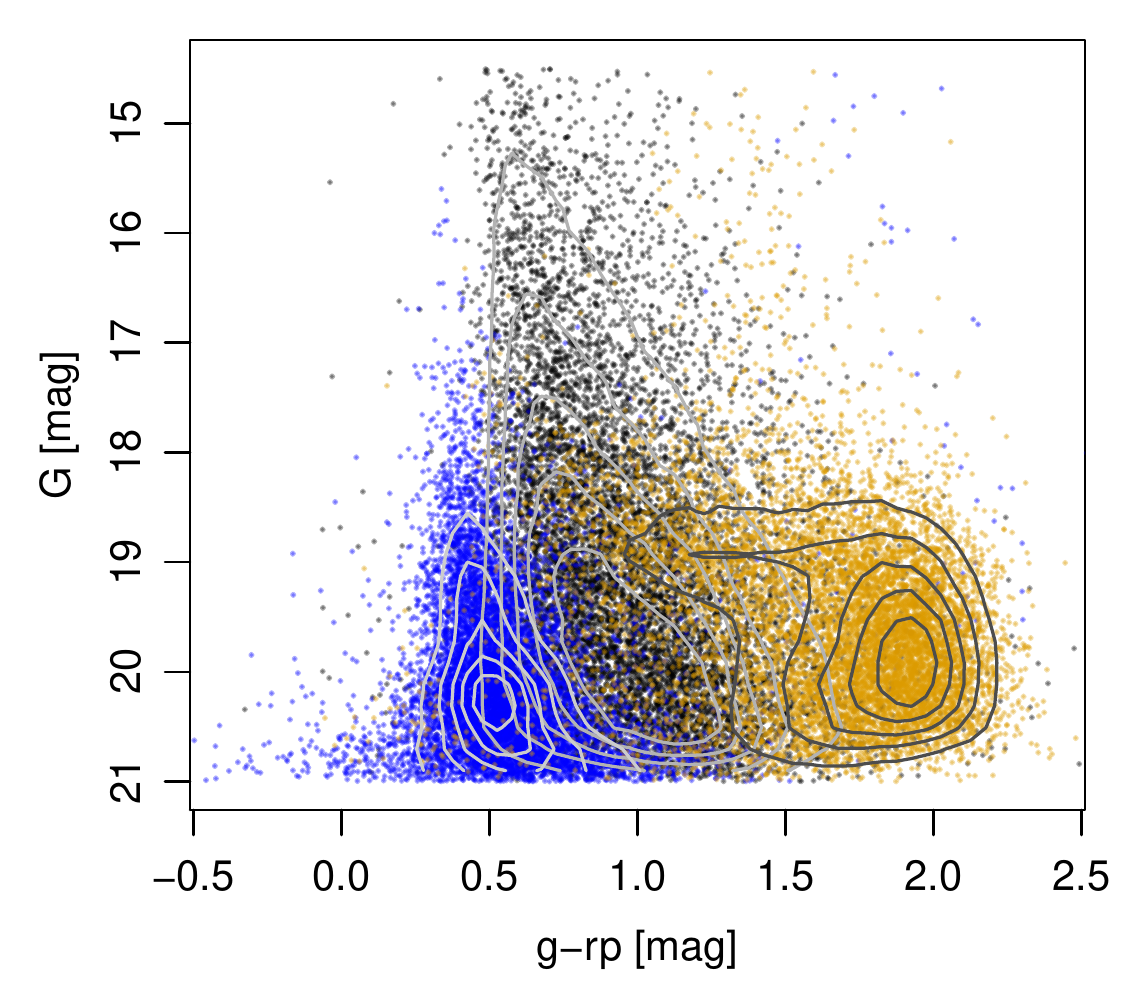}
\includegraphics[width=0.50\textwidth, angle=0]{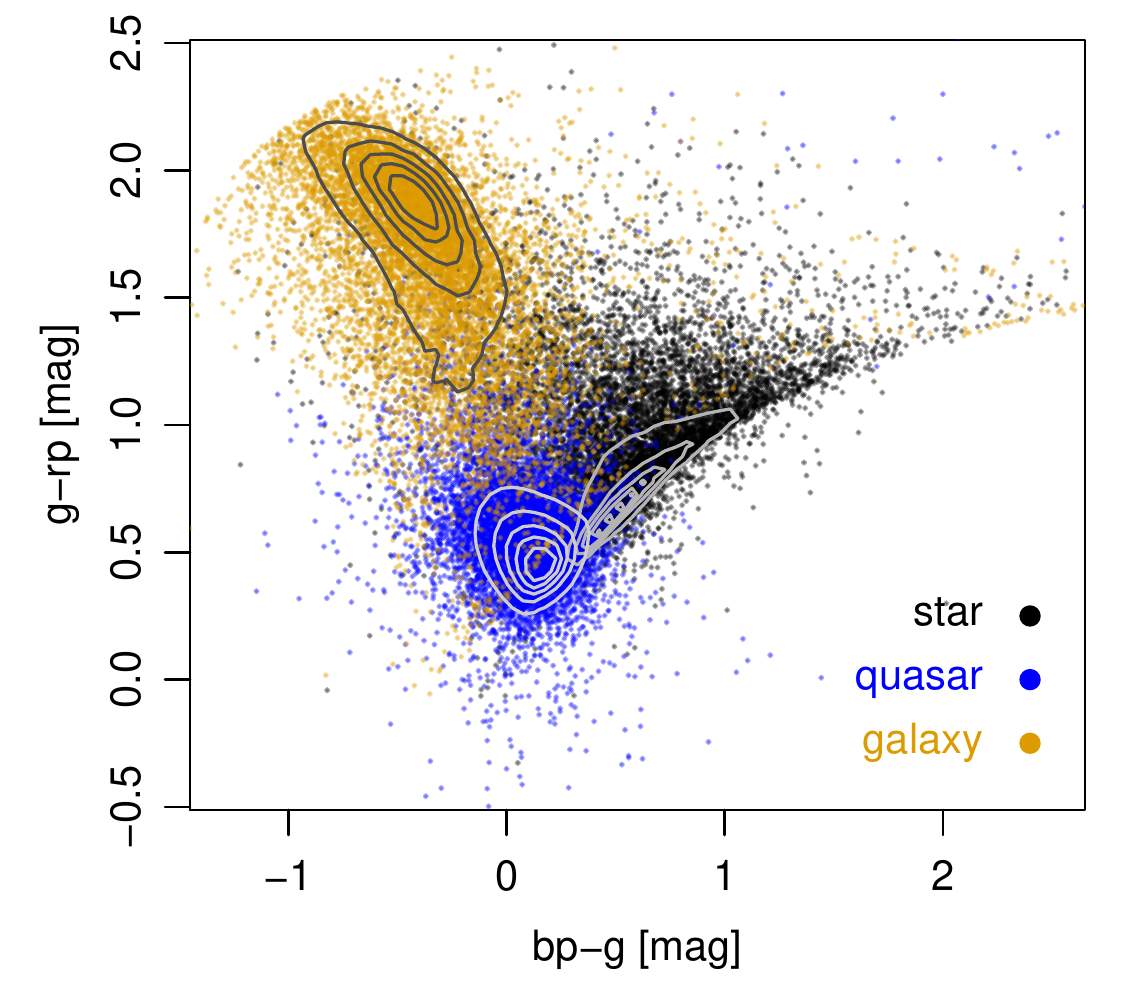}
\caption{Colour--magnitude diagram (top) and colour--colour diagram (bottom) for objects in the application set assigned to classes when the posterior probability is greater than 0.5. The colour indicates the assigned class, black: stars, blue: quasars, orange: galaxies (plotted in that order).
  10\,000 objects have been selected at random from each class.
   The contours for each class (built from all the data in each case) show equal spaced densities from
   $1/6$ to $5/6$ of the maximum density for that class.
  The classes overlap, so we have plotted them separately in Figure~\ref{fig:apply_any_0p5threshold_threepanels_onecmd} in the appendix.
\label{fig:apply_any_0p5threshold_onecmd}
}
\end{center}
\end{figure}

In contrast, the quasar and in particular the galaxy \gmag\-band magnitude distributions have peaks at fainter magnitudes in the application classes. This can be a result of both stellar contamination and the true objects in \gaia\ having a slightly fainter magnitude distribution than our (SDSS-based) training set. Recall Figure~\ref{fig:qso_redshift_mag_distributions}, which shows that the peak of the SDSS $g$-band magnitude distribution for the training set is at around 19.5.
The distributions in colour and magnitude are also visualized in Figure~\ref{fig:apply_any_0p5threshold_onecmd}, which can be compared to the corresponding diagrams for the true classes in the Figure~\ref{fig:any_plotsel_onecmd} (compare also Figure~\ref{fig:any_plotsel_threepanels_onecmd} with~\ref{fig:apply_any_0p5threshold_threepanels_onecmd}). Whereas the distributions for the quasars (and the stars) are quite similar, the galaxies in the application set show both a marked shift to fainter magnitudes and a change in the distribution for objects with $\grp\lesssim 1.5$. These regions are in the tails of the training set distribution, and so were perhaps modelled less well by the GMM.
The horizontal spur at $\gmag \simeq 19$\,mag stretching between $\grp \simeq 1$ and 1.5\,mag in the upper panel of Fig.~\ref{fig:apply_any_0p5threshold_onecmd} is predominantly stars in the Large and Small Magellanic clouds (LMC and SMC). 
We also see an excess of objects classified as galaxies along the colour edge (equation~\ref{eqn:colouredge}); these are predominantly stellar contaminants.

The latitude distribution in the applied classes shows that there is lower frequency of classified extragalactic objects at low latitudes. This is an expected consequence of dust absorption by the Galactic disk, which decreases the number of lines-of-sight through the Galaxy. The LMC/SMC and parts of the Galactic bulge are prominent peaks in the distributions for quasars and/or galaxies (explained in the next subsection).

We also some differences in the distributions for \relvarg. The galaxies show an excess of low amplitude variables in the application set. As the distribution for stars is ramping up towards low amplitudes, this is almost certainly a signal of the stellar contamination. Quasars show a more complicated change in \relvarg, in particular an excess of higher amplitude variables. As there are relatively few stars with such high amplitudes, and there is very little contamination of galaxies into the quasars (see Table~\ref{tab:cm_threshold}), this suggests that we are picking up a larger proportion of higher-amplitude variable quasars than we had in our training set.

Finally, the distribution of \uwe\ is largely unchanged for the quasars, but the galaxy class now shows a second peak just below 2, and no longer the broad hump at around 4. There is no clear-cut reason for these, although the loss of the hump into the star class would not be noticeable in the distribution for stars, as they are far more numerous.

\subsubsection{Distribution on the sky}\label{sec:sky_distribution}

\begin{figure*}
\begin{center}
\includegraphics[width=0.80\textwidth, angle=0]{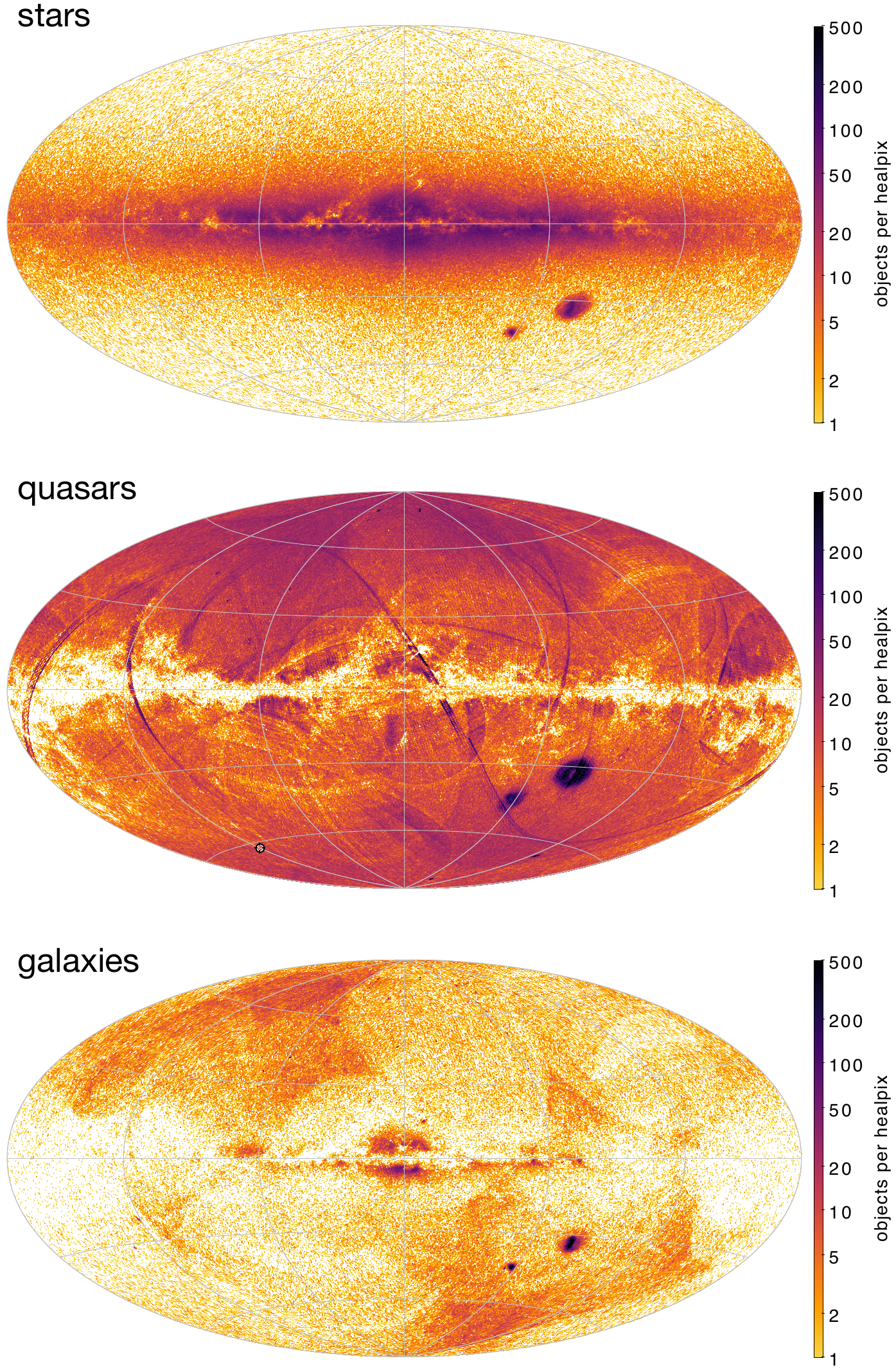}
\caption{Distribution in Galactic coordinates for objects in the application set assigned to classes when the posterior probability is greater than 0.5, shown separately for each class. Each plot shows the number of objects per healpix at level 7 (0.21 sq.\ deg.) on a logarithmic scale in an Aitoff (equal-area) projection.
The Galactic centre is in the middle and longitude increases to the left.
  The panels for quasars and galaxies show all objects with $P>0.5$, whereas the panel for stars is a random subset of one million objects with $P>0.5$ (so multiply the scale for stars by about $10^3$ to get the true numbers per healpix).
\label{fig:apply_0p5threshold_galsky}
}
\end{center}
\end{figure*}

Figure~\ref{fig:apply_0p5threshold_galsky} shows the distribution on the sky of the $P>0.5$ sample for each assigned class. 
The spatial distribution for stars follows our expectations, with a concentration of objects towards the Galatic plane and bulge.

The LMC and SMC stand out in all three panels.  Wherever there is a high density of objects on the sky, like in the LMC/SMC, the density of misclassified objects will likewise be high, even if the misclassification rate is no higher.  In addition to this, the LMC and SMC are so distant that the parallaxes and proper motions of their constituent stars are statistically insignificant, like many extragalactic objects, so we may also have a higher misclassification rate.

The distribution for quasars displays two features. First, there is a lower density at low latitudes that appears to follow the dust in the Milky Way. This complex pattern is not learned from the training data, because \sinb\ was randomized for extragalactic objects in the training, and longitude is not used at all.  The fact that we don't find quasars in these regions is presumably because they are genuinely absent from the \gaia\ catalogue due to dust obscuration. The second feature is the presence of numerous stripes. These are a consequence of how \gaia\ scans the sky (on quasi great circles) and the fact that it performs on-board object detection independently each time it crosses the same point on the sky: Objects near the magnitude detection limit and/or in crowded regions may be recorded different numbers of times in each scan. This determines -- together with various quality thresholds and filtering done during processing -- whether or not an object has sufficient data to define all eight features required for our classification. Just as with the LMC/SMC, a variable number of objects produces a variable number of misclassifications, which is why the stripes are different for the three classes in Figure~\ref{fig:apply_0p5threshold_galsky}.
(They are much less visible in the stars' plot because the misclassification rate is so low.)
There are many factors involved in determining this, so there is a not a simple mapping between the scanning law and the stripes we see. Further discussion and examples can be found in \cite{2018A&A...616A..17A}.

\subsubsection{Distribution in WISE colours}

\begin{figure}
\begin{center}
\includegraphics[width=0.50\textwidth, angle=0]{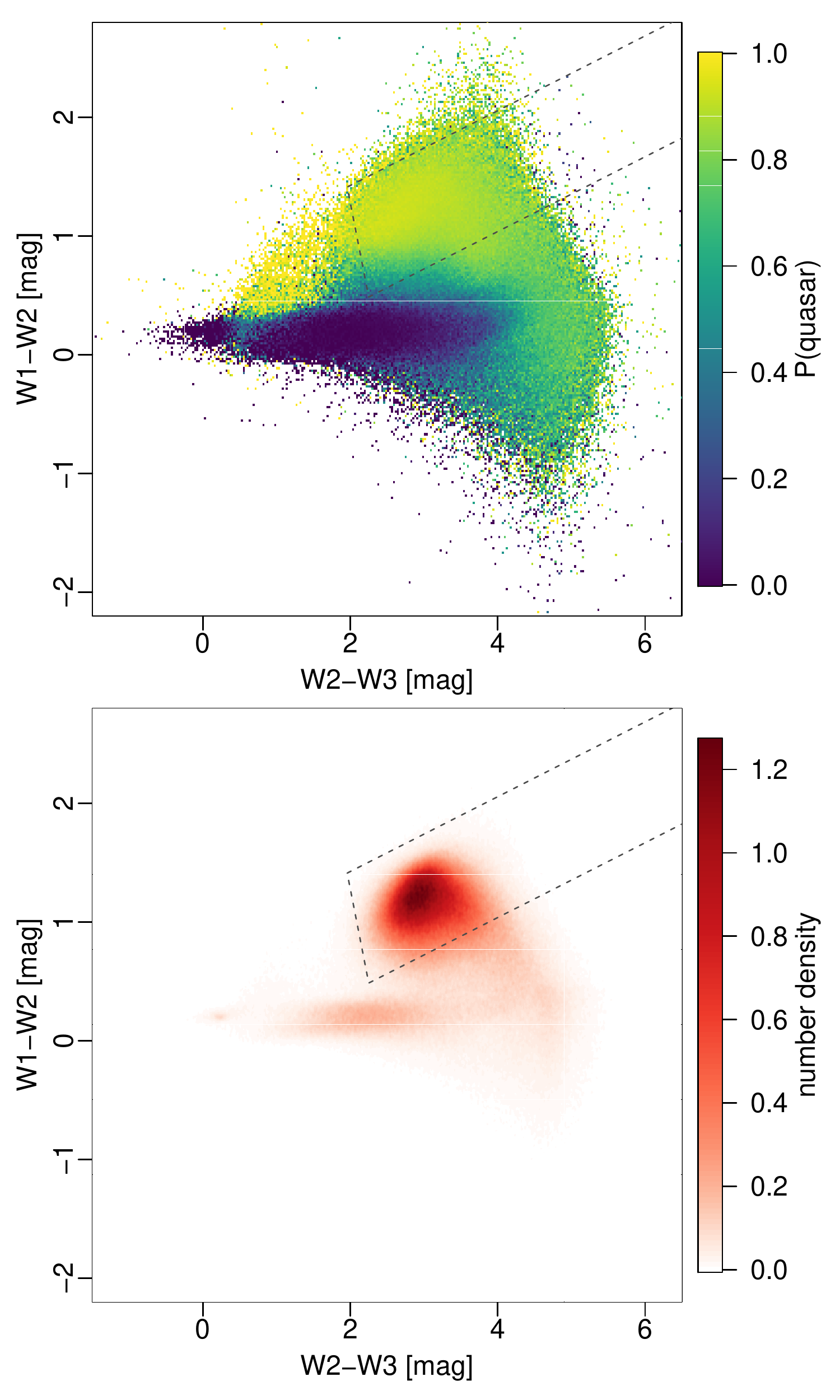}
\caption{WISE colours of objects in our extragalactic sample, colour coded by the mean \pqso\ in small colour bins (upper panel), or by the density of objects as computed by a kernel density estimate (lower panel; the colour bar covers the full range of densities).
  Of the 2.7 million objects in this sample, 1.1 million 
  have all three WISE colours and so are shown here.
  The grey dashed polygon indicates the wedge defined by \citet{2012MNRAS.426.3271M} to identify AGN using these WISE colours.
\label{fig:apply_all_WISE_pqso}
}
\end{center}
\end{figure}

Figure~\ref{fig:apply_all_WISE_pqso} shows the distribution of objects in our extragalactic sample in a WISE colour space.  The bands W1, W2, and W3 are located at 3.4\,\um, 4.6\,\um, and 12\,\um\ respectively.  1.1 million objects in our sample (42\%) have measurements in all three of these bands.  We overplot the ``wedge'' defined by \cite{2012MNRAS.426.3271M} to identify AGN.  This wedge has been used by various authors, for example \cite{2018ApJS..236...37P} and \cite{2018A&A...616A..14G} who identified around half a million AGN in \gdr{1} and \gdr{2} respectively (see introduction). We see from Figure~\ref{fig:apply_all_WISE_pqso} that objects in our sample with the highest value of \pqso\ are found in this wedge. 66\% of our extragalactic sample 
falls inside the wedge, with a median value of \pqso\ of 0.97 (the interquartile range is 0.823--0.998).
The near horizontal slab of low \pqso\ objects at W1$-$W2\,$\simeq$\,0\,mag comprises high probability galaxies plus stellar contaminants. In the upper panel of Figure~\ref{fig:apply_all_WISE_pqso} we see a second clump of objects with high \pqso\ at W2$-$W3\,$<$\,2\,mag, W1$-$W2\,$>$\,0\,mag. Most of these have $\bpg>0.5$ and $\grp>1$ and are therefore outside the main set of quasars in the training set (cf.\ Fig.~\ref{fig:any_plotsel_threepanels_onecmd}). This classification of these objects as quasars is therefore an extrapolation of the GMM.  This observation, together with the fact that most objects in this second clump are in dense regions on the sky (the Galactic plane or the LMC/SMC) and so may have spurious colours, suggests they are misclassifications. They are quite rare, however (just 3000 of the 1.1 million with WISE colours).
The set of objects in Figure~\ref{fig:apply_all_WISE_pqso} with W2$-$W3\,$>$\,4\,mag (and outside the wedge)
with reasonably high \pqso\ do, in contrast, have a \bpg, \grp\ colour distribution similar to the quasars in the training set, suggesting that many of them are indeed quasars (and they all lie outside of dense sky regions).
A comparison with Figure 5 of \cite{2012MNRAS.426.3271M} indicates that some of these could be objects for which the radiation emitted in these bands is dominated by the host galaxy.

\subsubsection{Distribution in redshift}\label{sec:quasar_redshifts}

\begin{figure}
\begin{center}
\includegraphics[width=0.50\textwidth, angle=0]{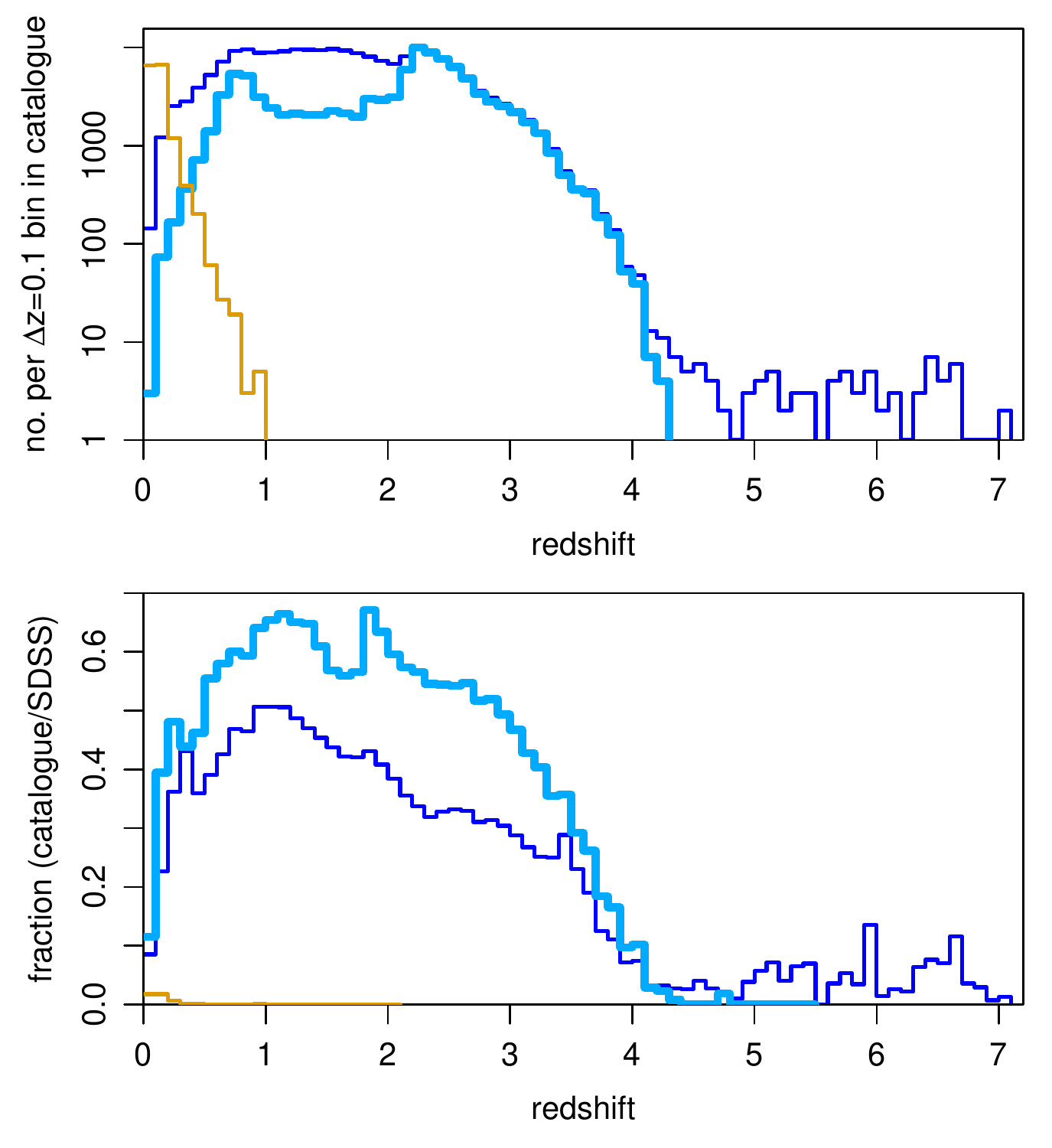}
\caption{Redshift distribution in the extragalactic sample for objects with $\pqso>0.5$ (blue) and for objects with $\pgal>0.5$ (orange). The thick light-blue line is for the subset of quasars with visually-verified redshifts. Upper panel: Number of objects. Lower panel: Number of objects as a fraction of all SDSS15 objects of that class (i.e.\ of SDSS class {\tt QSO} for the blue line and SDSS class {\tt GALAXY} for the orange line).
\label{fig:apply_qso_redshifts_logdensity}
}
\end{center}
\end{figure}

We obtain redshifts for our extragalactic sample from the SDSS-DR15 spectroscopic catalogue.  We select all objects with non-negative redshifts that are class {\tt QSO} or {\tt GALAXY}, and that meet the additional selection criteria specified in sections~\ref{sec:data_train_quasars} and~\ref{sec:data_train_galaxies}.
This gave 545\,632 objects of class {\tt QSO} and 2\,311\,066 of class {\tt GALAXY}.  A positional cross-match of these with a 1\arcsec\ radius to our extragalactic sample left 212\,592 objects of class {\tt QSO} with $\pqso>0.5$ (``quasars''), and 15\,205 objects of class {\tt GALAXY} with $\pgal>0.5$ (``galaxies'').  (The explanation for the big difference in fraction matched between the two classes was given in section~\ref{sec:data_train_galaxies}.) This is 9\% and 4\% respectively of our extragalactic sample. The distribution of these redshifts is shown in the upper panel of Figure~\ref{fig:apply_qso_redshifts_logdensity}.  The quasars extend to much higher redshifts than the galaxies. 3.9\% of the quasars in our extragalactic sample (that have redshifts) have $z>3$, and 0.08\% (165 objects) have $z>4$. The shape of this distribution does not change much if we limit it to objects with $\pqso>0.8$ or 0.9. Note, however,
that if we limit the sample to only those sources that have visually-verified redshifts (those with a defined value of {\tt z\_vi} in SDSS, shown with the thick light-blue line), we lose the highest redshift objects.
The lower panel shows these distributions expressed as a fraction of the number in SDSS-DR15 (for each redshift bin). The combination of the \gaia\ on-board detection together with our feature selection retains so few galaxies that the fraction for galaxies is virtually zero. For quasars, we find over 40\% of quasars in SDSS-DR15 with redshifts between 0.2 and 2.0 (higher for the visually-verified subset), with the fraction dropping to higher and lower redshifts.

\subsubsection{Estimating the true class fraction from the posterior probabilities}\label{sec:interpreting_probabilities}

We would like to estimate the true fraction of quasars and galaxies among the 1.20 billion objects in the application set. This is not straight forward due to the incompleteness and contamination, and there is more than one way to do it.

One approach is to use the posterior probabilities.  If a set of $N$ objects have posterior probabilities for class $k$ of $\{P_i\}$, then the expected number of objects that are truly of class $k$ is $\sum_{i=1}^N P_i$. For the complete set of objects in the application set this gives $\expnqso=$\,3\,502\,040 and $\expngal=$\,505\,367 (rounded to the nearest integer). These correspond to class fractions of  $29\times 10^{-4}$ for quasars and $4.2 \times 10^{-4}$ for galaxies.
This is similar to the class prior for quasars ($20\times 10^{-4}$), but three times larger than its value for galaxies ($1.3\times 10^{-4}$). This may not be the best way to infer the true class fractions, however, because $\sum_{i=1}^N P_i$ is dominated by
many objects with very small probabilities. Such probabilities may be poorly defined, because they are often obtained from the tails of the Gaussian distributions that are less well constrained during the training. The probabilities also do not take into account the measurement uncertainties on individual objects, and there are inevitably outright misclassifications. 

Two alternative ways of inferring the true class fractions that use the confusion matrix on the test set are explained in appendix~\ref{sec:expected_class_fraction}.
These produce values consistent with each other, but that are five times lower than those obtained above.
Because these only focus on the well-classified objects, we think these are more reliable. Our final inference of the true class fractions for quasars and galaxies (as defined by our training sets), for the 1.20 billion \gaia\ sources with all eight features and $\gmag>14.5$\,mag, are $5.7\times 10^{-4}$ and $0.91\times 10^{-4}$ respectively, each with a 1$\sigma$ uncertainty of about 50\% (see Table~\ref{tab:class_fractions}). These correspond to 690\,000 quasars and 110\,000 galaxies.
The former is consistent with the number of quasars that other authors have found in \gdr{2} by various means (see section~\ref{sec:introduction}).

\section{The catalogue} \label{sec:catalogue}

We provide the results from the application of our model to \gdr{2} (described in section~\ref{sec:results_application}) as a catalogue with three columns\footnote{\url{http://www.mpia.de/homes/calj/gdr2_class.html}}: the \gdr{2} source identifier (a long integer), \pqso, and \pgal. The two probabilities are rounded off to the sixth decimal place.  The star probability can be computed from $\pstar = 1-(\pqso+\pgal)$, and all other data obtained by cross-matching to \gdr{2} using the source identifier.  As our main goal is to identify extragalactic objects, we only report in this catalogue the extragalactic sample, i.e.\ those objects with $\pext=\pqso+\pgal>0.5$.
There are 2\,690\,021 such objects, a fraction 0.00224 of the 1.20 billion objects in \gdr{2} with all 8 features and $\gmag \geq 14.5$\,mag.
The catalogue is sorted by increasing \verb+source_id+.

The probabilities we provide assume the class prior of (star, quasar, galaxy)\,=\,$(7500, 15, 1)$. To adopt a different prior, simply divide the probability vector for each source by this class prior, multiply by the new class prior, and renormalize.

This set of quasars and galaxies with $\pext>0.5$ is a superset of the set obtained by
selecting by maximum probability. To see this, consider the necessary (and sufficient) conditions for assigning an object to the quasar class using maximum probability:
\begin{subequations}
\label{eqn:maxprob_conditions}
\begin{alignat}{2}
  \pqso \,>&\ \pgal \\
  \pqso + \frac{1}{2}\pgal \,>&\  \frac{1}{2} 
\end{alignat}
\end{subequations}
the latter coming from $\pqso > \pstar$.  If this second condition is true then so is $\pqso + \pgal > 1/2$, the condition to be in the catalogue.  The same argument applies for quasars when swapping \pqso\ and \pgal.  That the $\pext>0.5$ set can contain more objects than those obtained from maximum probability can be illustrated with the example $(\pstar, \pqso, \pgal)=(0.44, 0.30, 0.26)$. This object would be in the $\pext > 0.5$ set, but would not be classified as either quasar or galaxy using maximum probability.  The two conditions in equation~\ref{eqn:maxprob_conditions} can be used to find those objects in the catalogue that are classified as quasars by maximum probability (and the set of galaxies is found by swapping \pqso\ and \pgal\ in the equations).

\begin{figure}
\begin{center}
\includegraphics[width=0.50\textwidth, angle=0]{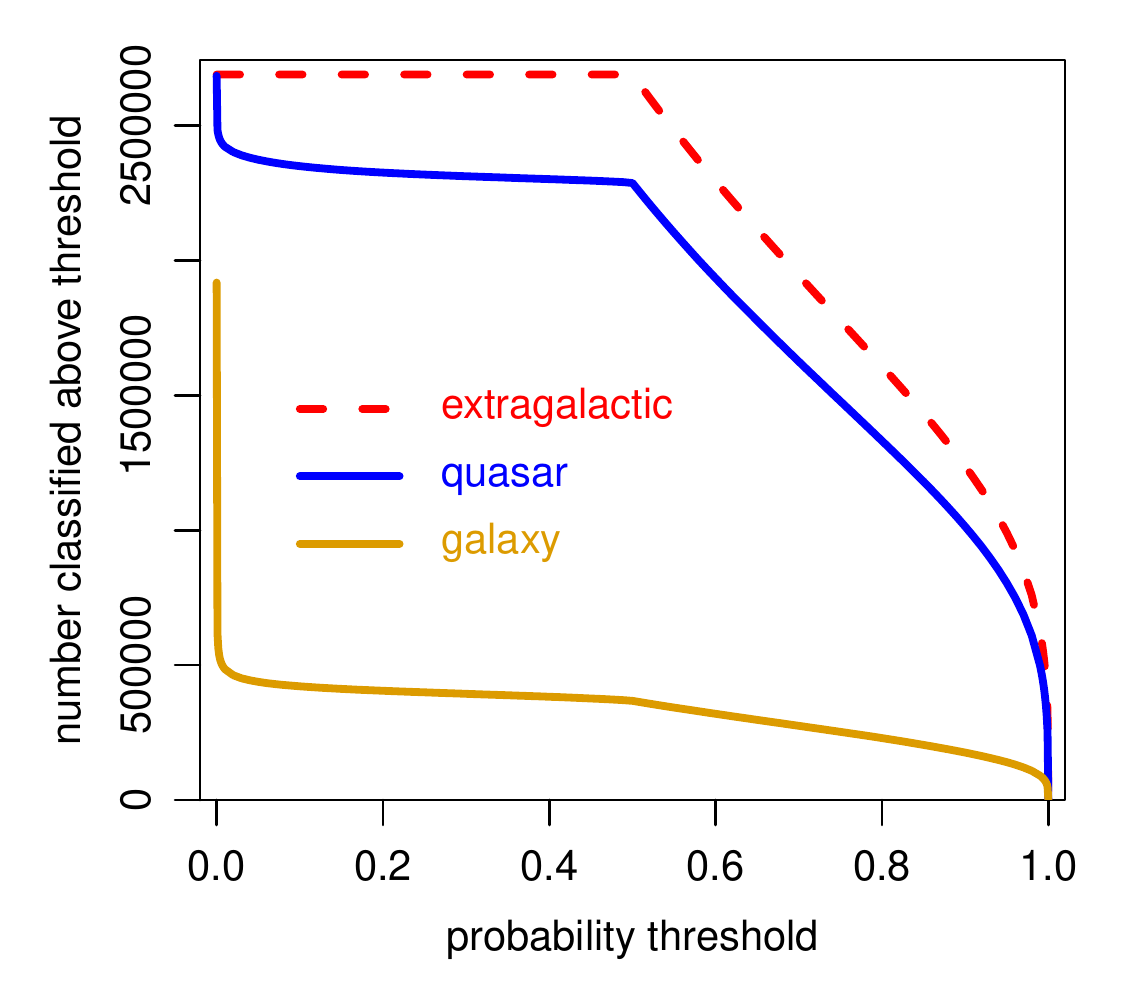}
\caption{The number of objects classified above a probability threshold in the catalogue for
  quasars (blue solid line) and galaxies (orange solid line) and for these two classes together (i.e.\ the sum of their posterior probabilities; red dashed line).
\label{fig:apply_qsogalext_nclass}
}
\end{center}
\end{figure}

The number of objects of each class in the catalogue with probabilities above a specified probability threshold are shown in Figure~\ref{fig:apply_qsogalext_nclass}.  The discontinuities at probabilities of 0.5 are a consequence of $\pext>0.5$.

The galaxy subset of our catalogue has considerable contamination, so should be used with caution.
The quasar subset is also contaminated, but its higher probability quasar members that are absent in other catalogues could be targets for spectroscopic follow-up. They may also be useful for future realizations of astrometric reference frames. Our catalogue extends across the Galactic plane, in contrast to many other searches for quasars. When accounting for the contamination and incompleteless, the catalogue as a whole can also be used to estimate the class prior for future classification projects (see section~\ref{sec:interpreting_probabilities}).

\section{Summary and conclusions}

We have used eight features based on photometry and astrometry from the second \gaia\ data release to identify quasars and galaxies. The purities of our samples are not particularly high according to our test set: 0.43 for quasars and 0.28 for galaxies (Table~\ref{tab:cm_maxprob}).
Purities of up to 0.60 can be achieved when adopting a high threshold, but at the price of a smaller sample (section~\ref{sec:apply_assigment_by_threshold}). This modest performance must nonetheless be seen in the context of the assumed rareness of the extragalactic objects (the class prior): 1 in 500 for quasars and 1 in 7500 for galaxies. The predicted completenesses are 0.58 for quasars and 0.73 for galaxies.

Our results are more satisfactory for quasars than they are for galaxies, not only in terms of sample purity, but also in terms of consistency with the training data and the nature of the contamination (sections~\ref{sec:results_apply_feature_space} and~\ref{sec:sky_distribution}).  The \gaia\ detection algorithm and subsequent data filtering severely limit the number of galaxies available. This results in a rather non-standard definition of our ``galaxy'' class, and yields only a small training sample.

To achieve better results we would either need a narrower definition of our classes, so as to reduce the overlap in the feature space with stars (although this would make them even rarer), or we need features that better discriminate between the classes, e.g.\ spectra. The GMM classifier itself performs well: We would achieve purities of around 0.95 if the objects in the three classes were equally common (section~\ref{sec:changing_prior}).

Our classification approach is purely empirical, so both avoids the need to model the \gaia\ selection function
and eliminates having to determine precise astrometric and photometric zeropoints and uncertainties.  The resulting distribution of classified objects in feature space, e.g.\ in Galactic sky coordinates, of course still shows the signature of the \gaia\ selection function (in particular the magnitude limit and scanning law).  As with any supervised machine learning method, our three classes -- star, quasar, galaxy -- are defined by the set of objects in their training sets.

We have shown that it is essential to incorporate an approximately-correct class prior into the classification process.  This influences both the classification probabilities and the (effective) class fractions in the test set.  If one instead assumes equal class priors (implicitly or explicitly), then the predicted purity of the rarer class(es) of objects will be hugely overestimated: Both the predicted purity would be higher -- at around 0.95 for both quasars and galaxies -- compared to the prediction using the correct class prior, and the actual purity considerably lower -- 0.050 for quasars and 0.0052 for galaxies -- compared to the model with the correct class prior (section~\ref{sec:changing_prior}).

Among the 1.20 billion objects in \gdr{2} that have $\gmag \geq 14.5$ and the eight features we require, we identify 2.3 million objects as quasars and 0.37 million objects as galaxies, with individual posterior probabilities of 0.5 or more.
We provide the classification posterior probabilities so that samples with a determinable trade-off between completeness and purity can be selected. The number of objects classified of course depends on the class prior to some degree (section~\ref{sec:changing_prior}). Our probabilities can easily be adjusted to reflect a different prior from the one we adopted (section~\ref{sec:catalogue}).
Using the above measured numbers of objects together with the confusion matrix on the test data, we can infer the true number of extragalactic objects in the \gdr{2} sample (section~\ref{sec:interpreting_probabilities}). We estimate 690\,000 quasars and 110\,000 galaxies, which correspond to class fractions of $5.7\times 10^{-4}$ and $0.91\times 10^{-4}$ respectively (the 1$\sigma$ accuracy in these figures is about 50\%).  There are far fewer galaxies than quasars because the \gaia\ observing mode and data processing selects strongly against extended objects.

Better classifications for many objects could be obtained using additional non-\gaia\ data, such as additional optical and infrared photometry, although this is not available homogeneously for the 1.20 billion objects we classify here.  Spectra would enable even better classifications, of course. We are expecting of order $10^8$ low resolution optical (BP/RP) spectra to be released as part of the third \gaia\ data release, planned for 2021. We will use these spectra, supported by a classifier similar to the one described here, to provide object classifications in that data release. Use of these spectra should permit a better separation of the classes and furnish more reliable probabilities, and therefore allow us to construct purer samples than could be attained in the present work.

\section*{Acknowledgements}

We thank Bernd Husemann, Jan-Torge Schindler, Ron Drimmel, Ludovic Delchambre, Eduardo Ba\~nados, Jan Rybizki, and the referee for useful comments on this work.  We are grateful to Mark Taylor for the TOPCAT software \citep{2005ASPC..347...29T} which we used extensively in this work. The data processing and most plots were done in R (\url{http://www.r-project.org}).
This study used data from the European Space Agency (ESA) mission \gaia\ (\url{http://www.cosmos.esa.int/gaia}), processed by the \gaia\ Data Processing and Analysis Consortium (DPAC, \url{http://www.cosmos.esa.int/web/gaia/dpac}). Funding for the DPAC has been provided by national institutions, in particular the institutions participating in the \gaia\ Multilateral Agreement.  This work was funded in part by the DLR (German space agency) via grant 50 QG 1403.

\bibliographystyle{mnras}
\bibliography{gdr2_class}

\appendix

\section{Distribution of the training features}


\begin{figure*}
\begin{center}
\includegraphics[width=1.0\textwidth, angle=0]{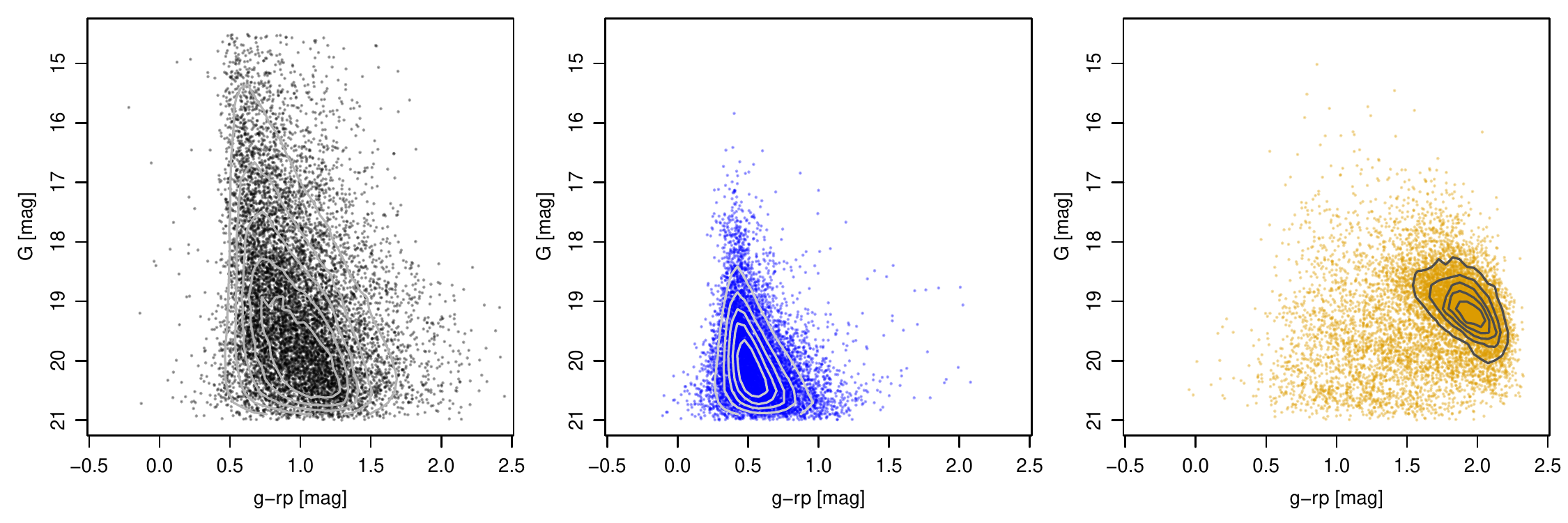}
\includegraphics[width=1.0\textwidth, angle=0]{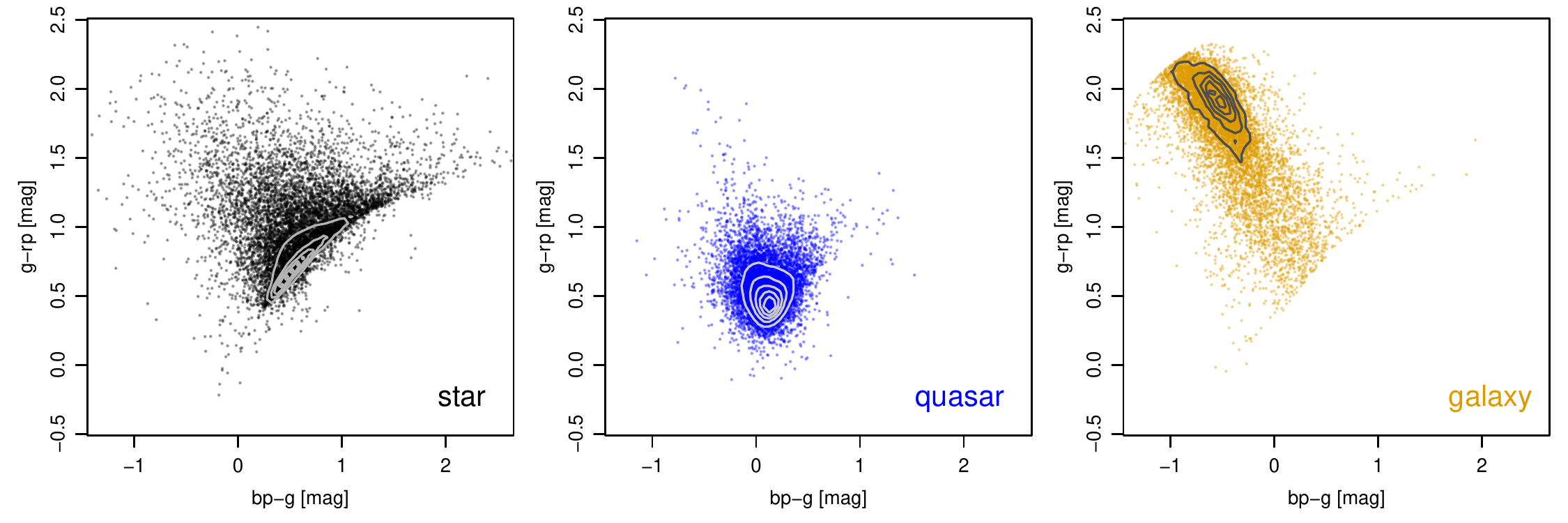}
\caption{Colour--magnitude diagrams (top) and colour--colour diagrams (bottom) for the training set plotted in separate columns according to their true classes. From left to right: stars, quasars, galaxies.
    The contours for each class (built from the entire train/test data set in each case) show equal spaced densities from
    $1/6$ to $5/6$ of the maximum density for that class.
  This is the same plot as Figure~\ref{fig:any_plotsel_onecmd} but with each true class in its own panel.
\label{fig:any_plotsel_threepanels_onecmd}
}
\end{center}
\end{figure*}

\begin{figure*}
\begin{center}
\includegraphics[width=1.0\textwidth, angle=0]{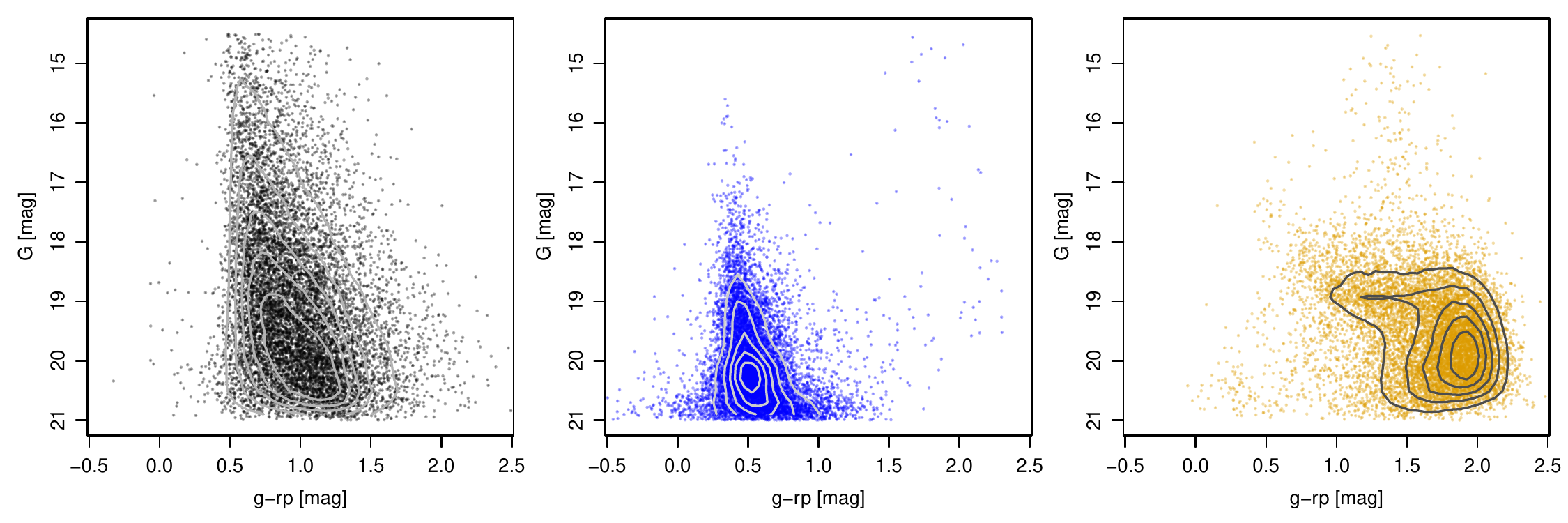}
\includegraphics[width=1.0\textwidth, angle=0]{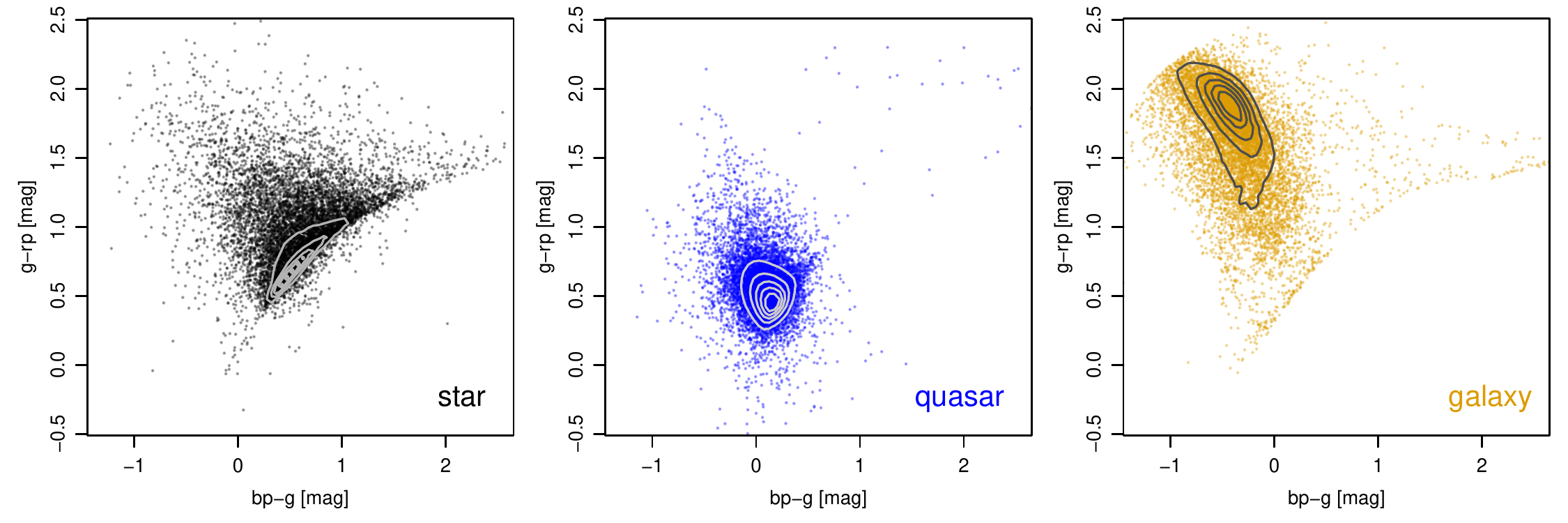}
\caption{Colour--magnitude diagrams (top) and colour--colour diagrams (bottom) for objects in the application set assigned to classes (columns of this plot) when the posterior probability is greater than 0.5.
From left to right: stars, quasars, galaxies.
   The contours for each class (built from all the data in each case) show equal spaced densities from
   $1/6$ to $5/6$ of the maximum density for that class.
  This is the same plot as Figure~\ref{fig:apply_any_0p5threshold_onecmd} but with each assigned class in its own panel.
\label{fig:apply_any_0p5threshold_threepanels_onecmd}
}
\end{center}
\end{figure*}

Figures~\ref{fig:any_plotsel_threepanels_onecmd} and~\ref{fig:apply_any_0p5threshold_threepanels_onecmd} are Figures~\ref{fig:any_plotsel_onecmd}
and~\ref{fig:apply_any_0p5threshold_onecmd} respectively,
but with each class shown in a separate panel.

\section{Estimating the true class fraction from the measured class fraction and the confusion matrix}
\label{sec:expected_class_fraction}

\subsection{Inversion}

We want to estimate the true fraction of objects $t_k$ in each class $k$.  Let $m_k$ be the measured fraction of objects classified into class $k$ (using maximum probability).  The measured fraction differs from the true fraction because of incompleteness and because of contamination from other classes.  Let $c_{ij}$ be the fraction of true class $i$ that is classified into measured class $j$. When $i=j$ this is the completeness of class $j$, and when $i \neq j$ this is the fraction of true class $i$ that contaminates measured class $j$.  It follows that \begin{equation}
  m_j \,=\, c_{jj}t_j + \sum_{i \neq j} c_{ij}t_i \ .
  \label{eqn:mf_tf_basic}
\end{equation}
The $c_{ij}$ are the elements ($i$th row, $j$th column) of the row-normalized (rows summing to one) confusion matrix.
For a $K$-class problem, equation \ref{eqn:mf_tf_basic} gives a system of $K$ linear equations in $K$ unknowns, which can be written in matrix form as
  \begin{equation}
    \mvec \,=\, \Cvec\trans \tvec
    \label{eqn:mf_tf_matrix}
  \end{equation}
  where $\mvec = \{m_j\}$ and $\tvec = \{t_i\}$ are $K$-element column vectors. $\Cvec\trans$ is the transpose of the row-normalized confusion matrix and may be considered as a transition matrix.  Equation~\ref{eqn:mf_tf_matrix} can be solved by matrix inversion to give an estimate of the true class fraction \tvec.  Applying this to the measured class fractions
found in section~\ref{sec:results_application}, along with the confusion matrix in Table~\ref{tab:cm_maxprob}, the estimate for the true class fractions is $5.8 \times 10^{-4}$ for quasars and $0.87 \times 10^{-4}$ for galaxies.

A difficulty here is that the confusion matrix is noisy, especially for galaxies, and this noise can be amplified and introduce asymmetry via the nonlinear operation of inverting $\Cvec\trans$. Another potential problem is that
the elements of \tvec\ are not constrained to be positive (although because \mvec\ is normalized, equation~\ref{eqn:mf_tf_matrix} ensures \tvec\ is too).

\subsection{Trinomial likelihood}

\begin{table}
\begin{center}
  \caption{The raw confusion matrix
(i.e.\ before multiplying each row by $\lambda_i$ from equation~\ref{eqn:testmodfac}) corresponding to Table~\ref{tab:cm_maxprob}.
\label{tab:cm_maxprob_raw}
}
\begin{tabular}{rrrr}
\hline
                     & star & quasar & galaxy \\
  \hline
  STAR     & 99818 &  157      &    25 \\
 QUASAR & 41735  &  58114   & 151  \\
 GALAXY & 2083     & 99       & 5818 \\
\hline
\end{tabular}
\end{center}
\end{table}

A more robust estimate of \tvec\ can be obtained using total regression.
We consider \Cvec\ as a measured, noisy version
of the true, unknown confusion matrix $\tilde\Cvec$. If we had this, then
the measured class fraction would be $\tilde{\mvec}$, i.e.\ $\tilde{\mvec} = \tilde\Cvec\trans \tvec$ (and equation~\ref{eqn:mf_tf_matrix} is an approximation which we now don't use).
We model the number of objects classified into the three classes in the application data set -- call these $N_1$, $N_2$, and $N_3$ -- with a trinomial likelihood that depends on $\tilde{\mvec}$, i.e.\
\begin{equation}
P(N_1, N_2, N_3 \given \tilde{\mvec}, \sum_k N_k) \propto {\tilde m}_1^{N_1} {\tilde m}_2^{N_2} {\tilde m}_3^{N_3} \ .
\end{equation}
We can write down a similar trinomial likelihood for the elements in the first
row of the (unnormalized) confusion matrix -- call these $\{n_{1j}\}$ -- in terms of the first row of $\tilde\Cvec$, and likewise for rows 2 and 3.
The joint likelihood for $\{N_k\}$ and $\{n_{ij}\}$ is then
\begin{alignat}{2}
{\cal L} \,=&\,   P \left( N_1,N_2,N_3\given \tilde{\mvec}=\tilde\Cvec\trans \tvec, \, \sum_k N_k \right) \nonumber \\
& \times  \prod_{i=1}^3 P \left( n_{i1},n_{i2},n_{i3} \given \tilde\Cvec_i, \, \sum_j n_{ij} \right)
\label{eqn:trinomial_likelihood}
\end{alignat}
where $\tilde\Cvec_i$ indicates the $i$th row of $\tilde\Cvec$.
Note that the numbers $\{n_{ij}\}$ are not the elements of the (modified) confusion matrix in Table~\ref{tab:cm_maxprob}, as those numbers have been adjusted to be appropriate for a test set with class fractions equal to the class prior. For the trinomial likelihood we must instead use the actual numbers in the confusion matrix prior to adjustment. These are given in Table~\ref{tab:cm_maxprob_raw}.

The likelihood in equation~\ref{eqn:trinomial_likelihood} has eight unknowns (three elements of \tvec, plus nine elements of $\tilde\Cvec$, minus four normalization constraints). The maximum of this has no analytic solution, so we instead adopt a uniform prior on the parameters and sample the resulting posterior using a Markov Chain Monte Carlo algorithm.
We summarize the resulting marginal distributions over each element of \tvec\ using the median and the central 68\% confidence interval.  These are reported in the bottom two lines of Table~\ref{tab:class_fractions} along with other estimates for comparison.

\begin{table}
\begin{center}
  \caption{Estimates of the class fractions of galaxies and quasars among the 1.20 billion sources in \gdr{2} that have all eight features.
    The first line is the class prior.
    The second is the fraction obtained when we assign classes according to maximum probability.  This estimate is affected by both the incompleteness and contamination of the classifier which we attempt to correct for in the next two lines.
    The third line gives the expected true fraction computed from the sum of the posterior probabilities (see section~\ref{sec:interpreting_probabilities}).
    The fourth line uses a direct inversion of equation~\ref{eqn:mf_tf_matrix} to estimate the true class fraction from this measured class fraction.
    The fifth line is a more robust estimate of this true class fraction by modelling the likelihood of both the measured class fraction and confusion matrix via the trinomial distribution; the final line expresses the uncertainty in this estimate.
\label{tab:class_fractions}
}
\begin{tabular}{rrr}
\hline
  Method       & quasars  & galaxies \\
\hline
  class prior  & $20\!\times\!10^{-4}$ &  $1.3\!\times\!10^{-4}$ \\
  measured  & $19\!\times\!10^{-4}$ & $3.1\!\times\!10^{-4}$ \\
  posterior probabilities  & $29\!\times\!10^{-4}$ & $4.2\!\times\!10^{-4}$ \\
  inversion   & $5.8\!\times\!10^{-4}$ & $0.87\!\times\!10^{-4}$ \\
  trinomial likelihood  & $5.7\!\times\!10^{-4}$ & $0.91\!\times\!10^{-4}$ \\
  (68\% confidence interval)       & 3.6--$7.8\!\times\!10^{-4}$ & 0.37--$1.5\!\times\!10^{-4}$ \\ 
\hline
\end{tabular}
\end{center}
\end{table}

The class fractions in \gdr{2} estimated in this way are similar to what we obtained from the simple inversion (line 4 of the table).  They are of course lower (by a factor of about 3.5) than the class fractions obtained from just counting the number of objects classified by maximum probability (line 2), because this latter estimate obviously includes contaminants. They are also lower than the expected true fractions we computed in section~\ref{sec:interpreting_probabilities} using the sum of the posterior probabilities (line 3); see that section for more discussion.

\end{document}